\documentclass[a4paper,aps,prd,10pt,preprintnumbers,showpacs,showkeys,twocolumn,superscriptaddress,nofootinbib]{revtex4-1}
\usepackage{orcidlink,graphicx,cmap,amsmath}
\usepackage[utf8]{inputenc}
\usepackage[T1]{fontenc}

\def\Order#1{{\cal O}\left(#1\right)}

\begin{document}

\title{Dark matter halo as a source of regular black-hole geometries}

\author{R. A. Konoplya \orcidlink{0000-0003-1343-9584}}
\email{roman.konoplya@gmail.com}
\affiliation{Research Centre for Theoretical Physics and Astrophysics, Institute of Physics, Silesian University in Opava,\\ Bezručovo náměstí 13, CZ-74601 Opava, Czech Republic}

\author{A. Zhidenko \orcidlink{0000-0001-6838-3309}}
\email{olexandr.zhydenko@ufabc.edu.br}
\affiliation{Centro de Matemática, Computação e Cognição (CMCC), Universidade Federal do ABC (UFABC),\\ Rua Abolição, CEP: 09210-180, Santo André, SP, Brazil}

\begin{abstract}
We construct exact black hole solutions free of curvature singularities, sourced by dark matter halos described by galactic density profiles. Regularity of the geometry is ensured by adopting the relation $P_{r}=-\rho$ between radial pressure and density, which is consistent with the phenomenological freedom of halo models. Under the assumptions of regularity and the weak-energy condition, sufficiently dense dark matter halos can give rise to asymptotically flat, singularity-free black holes embedded in a galactic environment. These regular black holes are shown to be stable under axial perturbations. In particular, we obtain solutions corresponding to Einasto and Dehnen-type dark matter profiles. We further compute the shadow radii and Lyapunov exponents associated with photon circular orbits around these black holes.
\end{abstract}


\maketitle

\section{Introduction}
The resolution of singularities predicted by classical general relativity remains one of the central challenges in black-hole physics. According to the singularity theorems, geodesic incompleteness is inevitable under broad physical assumptions, suggesting that classical black holes conceal regions where the known laws of physics break down. A wide variety of approaches have been proposed to resolve the central singularities of classical black holes.
One class of models is based on phenomenological regular metrics, where the spacetime geometry is modified at short distances in such a way that curvature invariants remain finite, as in the constructions by Bardeen, Hayward, Simpson and Visser~\cite{Bardeen:1968,Hayward:2005gi,Simpson:2018tsi}.
Another line of research employs effective matter sources, often interpreted as nonlinear electrodynamics or anisotropic fluids, which regularize the geometry while preserving asymptotic flatness~\cite{Ansoldi:2008jw,AyonBeato:1998ub,Dymnikova:1992ux,Bronnikov:2000vy,Bronnikov:2024izh,Bronnikov:2005gm}.
Regular black holes have also emerged in quantum-gravity-inspired frameworks, including loop quantum gravity, asymptotically safe gravity, quasi-topological and nonlocal or higher-derivative extensions of general relativity, where quantum corrections eliminate classical singularities~\cite{Kazakov:1993ha,Modesto:2008jz,Bonanno:2000ep,Lan:2023cvz,Bonanno:2023rzk,Bonanno:2025dry,Spina:2025wxb,Zhang:2024ney,Solodukhin:2025opw,Bueno:2024dgm,Bueno:2025tli,Bueno:2024eig,Frolov:2026rcm}. Various regular black-hole solutions can be obtained in brane-world scenarios \cite{Bronnikov:2006fu,Bronnikov:2003gx,Casadio:2001jg}.
More recently, systematic studies have examined the dynamical and radiative properties of regular black holes, including accretion, grey-body factors, Hawking radiation, and, most notably, quasinormal modes, revealing both qualitative similarities and quantitative deviations from their singular counterparts~\cite{Bronnikov:2012ch,Fernando:2012yw,Flachi:2012nv,Li:2013fka,Liu:2020ola,Jusufi:2020odz,Toshmatov:2015wga,Toshmatov:2018ell,Toshmatov:2019gxg,Rincon:2020cos,Yang:2021cvh,Franzin:2022iai,Konoplya:2022hll,Meng:2022oxg,Franzin:2023slm,Konoplya:2023ahd,Bolokhov:2023ruj,Konoplya:2023bpf,Skvortsova:2024eqi,Pedrotti:2024znu,Stashko:2024wuq,Skvortsova:2024wly,Khoo:2025qjc,Konoplya:2025uta,Bolokhov:2025egl,Bolokhov:2025lnt,Arbelaez:2025gwj}.

A key question in all such constructions is whether the required matter sources can be justified within astrophysical settings. Indeed, black holes in nature are never completely isolated but always reside in astrophysical environments dominated at large scales by dark matter. Overwhelming observational evidence supports the existence of galactic and cluster dark matter halos \cite{Navarro:1996gj,Bertone:2005xz}, although the microscopic nature of dark matter remains unknown. Phenomenological halo models, such as the Hernquist \cite{Hernquist:1990be}, Navarro–Frenk–White (NFW) \cite{Navarro:1996gj}, and other profiles, are widely used in galactic dynamics and cosmological simulations. Importantly, these models prescribe the density distribution without fixing a unique relation between pressure and density, thereby leaving freedom for effective fluid descriptions.

In this work we exploit this freedom to construct exact black hole solutions sourced by dark matter halos, focusing on the Einasto \cite{Einasto:1965czb} and Dehnen-type \cite{Dehnen:1993uh} density profiles. Regular black hole configurations associated with the Dehnen-type profile, which was referred to as the Dekel-Zhao profile \cite{Dekel:2017bwy,Zhao:1995cp}, were recently reviewed in \cite{Kar:2025phe}, with particular emphasis on solutions with infinite asymptotic mass. The barotropic subclass of such configurations was also discussed in \cite{Sajadi:2023ybm,Sajadi:2025prp}.

In the present study, we establish the general criteria and delineate the characteristic features of density profiles that give rise to regular black-hole geometries. We show that assuming the radial pressure satisfies $P_{r}=-\rho$ leads to spherically symmetric spacetimes that are everywhere regular and asymptotically flat, with horizons forming under suitable parameter conditions. Within this framework, we obtain a new family of exact solutions describing regular black holes immersed in galactic halos for a broad class of density profiles. In particular, we derive explicit and remarkably simple analytic asymptotically flat black-hole metrics for the Dehnen-type and Einasto distributions in specific parameter ranges. Since
the phenomenological profiles specify only the density, our choice of pressure does not conflict with halo phenomenology, while ensuring regularity of the interior. The resulting solutions can be interpreted as singularity-free black holes embedded in galactic environments. We will also show that the obtained configuration of the regular black hole and halo is stable against axial gravitational perturbations of spacetime. As a basic geometry probe of the obtained spacetime, we calculate the shadow radii for several representative examples of regular black holes.

The work is organized as follows. In Section~\ref{sec:basic} we formulate the general approach for constructing regular black-hole geometries sourced by spherically symmetric dark-matter distributions. Sections~\ref{sec:Einasto} and~\ref{sec:Dehnen} present explicit examples of such solutions, respectively, for the Einasto and Dehnen-type density profiles. The stability of the obtained configurations under axial perturbations is analyzed in Section~\ref{sec:axial}. The shadows cast by the obtained black holes and the Lyapunov exponents for null circular geodesics are considered in Section~\ref{sec:shadow}. In Section~\ref{sec:discussions} we discuss the physical interpretation and astrophysical relevance of these results, while the main conclusions are summarized in Section~\ref{sec:conclusions}.

\section{Spherically symmetric dark matter distributions and generic regular black hole solutions}\label{sec:basic}
Galactic matter is usually modeled by an anisotropic fluid with some density distribution, which implies an almost spherical halo dominated by dark matter \cite{Benson:2010de}. Depending on the size, mass, and form of a galaxy one or another distribution is preferable.

In order to simplify notations we employ the general form for the spherically symmetric line element,
\begin{equation}\label{line-element}
ds^2=-f(r)dt^2+\frac{B^2(r)}{f(r)}dr^2+r^2(d\theta^2+\sin^2\theta d\varphi^2),
\end{equation}
where we assume that $B(r)>0$.

It is convenient to introduce the mass function, \mbox{$m(r)\leq r/2$}, such that
\begin{equation}\label{massfunction}
1-\frac{2m(r)}{r}\equiv\frac{f(r)}{B^2(r)}.
\end{equation}

The horizon radius $r_0$ satisfies
\begin{equation}\label{horizoncond}
f(r_0)=0, \qquad r_0=2m(r_0).
\end{equation}

The external matter is the anisotropic fluid, so that the nonzero components of the corresponding stress-energy tensor are
\begin{equation}\label{stress-energy}
\begin{array}{rcl}
T_{t}^{t} &=&-8\pi\rho(r),\\
T_{r}^{r} &=& ~~8\pi P_r(r), \\
T_{\theta}^{\theta} = T_{\varphi}^{\varphi} &=& ~~8\pi P(r).
\end{array}
\end{equation}

After some algebra, the Einstein equations are reduced to the following form:
\begin{eqnarray}
\label{mder}
m'(r)&=&4\pi r^2\rho(r),\\
\label{Bder}
B'(r)&=&4\pi r^2B(r)\frac{\rho(r)+P_r(r)}{r-2m(r)}.
\end{eqnarray}

One can notice that, since $B(r_0)$ is finite, Eq.~(\ref{Bder}) implies the additional condition at the horizon
\begin{equation}\label{horzoncond}
\rho(r_0)+P_r(r_0)=0.
\end{equation}
In \cite{Konoplya:2022hbl} this condition for zero radial pressure, $P_r(r)=0$, yielded a modification in the density profile in such a way that $\rho(r_0)=0$.

In the present paper we do not modify the density profile of dark matter. Instead, we consider the vacuum equation of state \cite{Zaslavskii:2025oli},
\begin{equation}\label{vacuum}
P_r(r)=-\rho(r),
\end{equation}
which automatically satisfies Eq.~(\ref{horzoncond}).
While this condition is more naturally associated with dark energy than with dark matter on cosmological scales, several studies have explored fluids obeying $p=-\rho$, or approaching this limit in certain regimes, as possible unified descriptions of dark matter and dark energy~\cite{Kamenshchik:2001cp,Bento:2002ps,Scherrer:2004au,Dymnikova:2015yma}. Related concepts arise in models of vacuum-like or self-interacting dark matter, in which the dark component effectively behaves as a cosmological-constant term within overdense regions or compact objects~\cite{Balakin:2003tk,Sahni:2004ai}. It should be emphasized, however, that in our context this equation of state is assumed to hold only within a central part of a galaxy.

Then, it follows from Eq.~(\ref{Bder}) that
\begin{equation}\label{Bsol}
B(r)=1,
\end{equation}
and we find that the tangential pressure is given
\begin{equation}\label{Pex}
P(r)=-\frac{r}{2}\rho'(r)-\rho(r),
\end{equation}
satisfying the weak energy condition as long as $\rho(r)>0$ and $\rho'(r)<0$.

Finally, we can solve Eq.~(\ref{mder}),
\begin{equation}\label{msol}
m(r)=4\pi \intop_0^rx^2\rho(x)dx,
\end{equation}
where we assume that $m(0)=0$. In this case there is no singularity in the spacetime as long as the density is finite everywhere.

Thus, in the present paper, we consider density profiles that possess the following features:
\begin{enumerate}
    \item Monotonically decreasing and nonnegative:
    \begin{equation}\label{decreasingcond}
        \forall r\geq0:\quad \rho(r)\geq0, \quad \rho'(r)<0,
    \end{equation}
    which ensures that the weak energy condition is satisfied.
    \item Finite everywhere:
    \begin{equation}\label{finitecond}
        \forall r\geq0:\quad \rho(r)<\infty,
    \end{equation}
    which guarantees that the corresponding spacetime solution is free of singularities, in particular at $r=0$.
    \item Asymptotically convergent:
    We also assume that the integral in Eq.~(\ref{msol}) converges as $r\to\infty$,
    \begin{equation}\label{convergencecond}
        \intop_0^{\infty}x^2\rho(x)dx<\infty.
    \end{equation}
It should be noted that not all density profiles used to describe galactic halos satisfy this condition, since realistic galaxies have finite size and typically follow different distribution laws outside their boundaries. In this work, we do not consider such piecewise-defined profiles.
\end{enumerate}

\section{Einasto profile}\label{sec:Einasto}
An example of a density distribution obeying all the above conditions is the Einasto profile~\cite{Einasto:1965,Einasto:1989,Retana-Montenegro:2012dbd}:
\begin{equation}\label{density}
\rho(r)=\rho_0\exp\left[-\left(\frac{r}{h}\right)^{1/n}\right], \quad n>0,
\end{equation}
where $\rho_0$ is the density in the center, which is expressed in terms of the total mass
\begin{equation}\label{totalhalomass}
M=m(r\to\infty)=4\pi h^3 n\Gamma(3n)\rho_0.
\end{equation}

If we take the limit $h\to0$ while keeping the total mass (\ref{totalhalomass}) fixed, the density (\ref{density}) vanishes for $r>0$, and the solution reduces to the Schwarzschild metric with a singularity at $r=0$.

Having in mind a scenario in which supermassive galactic black holes could be formed during a long period of time as a result of merging of many smaller black holes, consideration of the Einasto profile does not sound completely exotic. Indeed, if some form of the Einasto profile were applicable in the early Universe, the galactic environment could be viewed as providing conditions conducive to the formation of a supermassive black hole. It is shown in \cite{Baes:2022pbc} that only the Einasto index $n\geq1/2$ can be supported by an isotropic orbital structure. The Einasto profile~\cite{Einasto:1965,Einasto:1989} has become one of the most successful empirical descriptions of dark-matter halos, accurately reproducing the smooth curvature of the density slope seen in dynamical simulations~\cite{Merritt:2006,MNRAS:Navarro2004,Springel:2008,Acharyya:2023rnq}. Subsequent works established its close connection to the Sérsic law for stellar systems and related the profile’s shape parameter to the halo’s mass-accretion history~\cite{Graham:2006,Ludlow:2013}.

It is worth emphasizing that the Einasto profile defines only the density and leaves the pressure profile undetermined. In particular, there is no intrinsic equation of state associated with the Einasto model, since the relation between pressure and density depends on the chosen dynamical description of the halo (e.g., collisionless particles with velocity dispersion governed by the Jeans equations, or an effective fluid approximation). Therefore, adopting the condition $P_{r}(r)=-\rho(r)$ for the radial pressure is not in conflict with the standard assumptions underlying the Einasto halo, but rather represents a phenomenological choice that ensures regularity of the black hole solution while remaining consistent with the absence of a fixed pressure–density relation in the halo model.

\begin{figure*}
\resizebox{\linewidth}{!}{\includegraphics{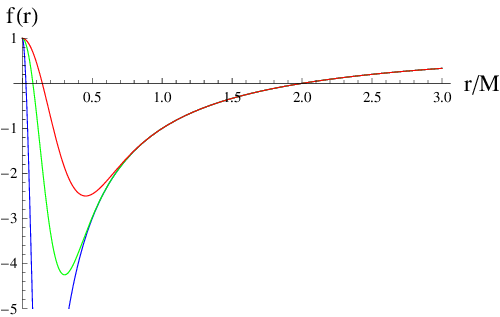}\includegraphics{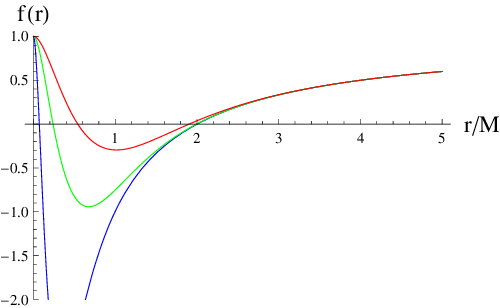}}
\caption{Metric functions for the black hole solutions with Einasto profile $n=1/2$ (left panel) and $n=1$ (right panel): $h=0.1M$ (blue), $h=0.2M$ (green), and $h=0.3M$ (red).}\label{fig:smallnmetric}
\end{figure*}

\begin{figure*}
\resizebox{\linewidth}{!}{\includegraphics{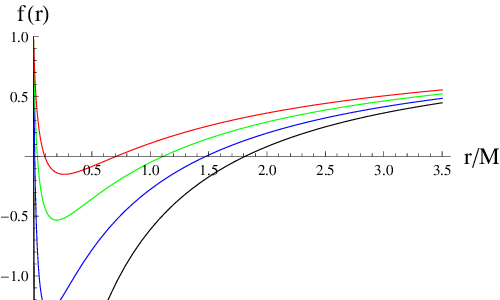}\includegraphics{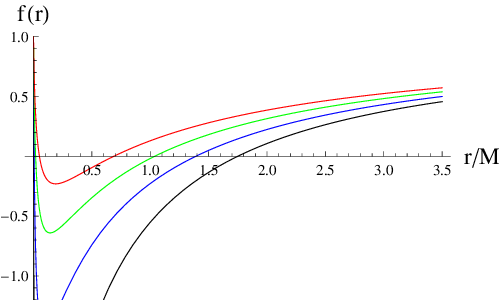}}
\caption{Metric functions for the black hole solutions with Einasto profile for large $n$.\\Left panel ($n=6$): $h=10^{-8}M$ (black), $h=2\cdot10^{-8}M$ (blue), $h=3\cdot10^{-8}M$ (green), and $h=4\cdot10^{-8}M$ (red).\\Right panel ($n=7$): $h=2\cdot10^{-10}M$ (black), $h=4\cdot10^{-10}M$ (blue), $h=6\cdot10^{-10}M$ (green), and $h=8\cdot10^{-10}M$ (red).}\label{fig:largenmetric}
\end{figure*}

Using the above general procedure, we now consider several examples of density profiles that lead to regular black hole solutions. In certain cases, these profiles admit simple analytic expressions, while in the general case the mass function must be obtained numerically by evaluating the corresponding integral. Both analytic and numerical examples are presented below. The accompanying \textit{Mathematica\textregistered} notebook\footnote{The \textit{Mathematica\textregistered} notebook is publicly available from \mbox{\url{https://arxiv.org/src/2511.03066v1/anc/BHEinasto.nb}}.} provides the code for generating solutions for arbitrary values of the profile parameters.

$\bullet$ For $n=1/2$ we have
\begin{equation}\label{nhmetric}
f(r)=1-\frac{2m(r)}{r}=1-\frac{2 M}{r} \text{erf}\left(\frac{r}{h}\right)+\frac{4 M e^{-r^2/h^2}}{\sqrt{\pi } h},
\end{equation}
where $\text{erf}(z)$ is defined as
$$\text{erf}(z)\equiv\frac{2}{\sqrt{\pi}}\intop_0^ze^{-x^2}dx,$$
and the asymptotic mass is given by
\begin{equation}\label{nhM}
M=\pi^{3/2}\rho_0h^3.
\end{equation}
The metric (\ref{nhmetric}) has the event horizon and inner horizon for sufficiently small $h\lesssim1.05M$.

The Ricci scalar is regular everywhere
\begin{equation}\label{Riccinh}
R=\frac{16 M e^{-r^2/h^2} \left(2 h^2-r^2\right)}{\sqrt{\pi } h^5},
\end{equation}
as well as the Kretschmann scalar, which has a cumbersome form, though for small $r$, we find:
$$R_{\mu\nu\lambda\sigma}R^{\mu\nu\lambda\sigma}=\frac{512 M^2}{3 \pi h^6}-\frac{512 M^2 r^2}{\pi h^8}+\Order{r}^4.$$

$\bullet$ For $n=1$ the metric takes the following form:
\begin{equation}\label{n1metric}
f(r)=1-\frac{2m(r)}{r}=1-\frac{2 M}{r}+M\frac{2 h^2+2 h r+r^2}{h^2 r}e^{-r/h},
\end{equation}
where
\begin{equation}\label{n1M}
M=8\pi\rho_0h^3.
\end{equation}

Again, the metric (\ref{n1metric}) has the event horizon and inner horizon for sufficiently small $h\lesssim0.388M$.
The Ricci scalar and Kretschmann scalar are regular
\begin{equation}\label{Riccin1}
R=Me^{-r/h}\frac{4 h-r}{h^4},
\end{equation}
$$R_{\mu\nu\lambda\sigma}R^{\mu\nu\lambda\sigma}=\frac{8 M^2}{3 h^6}-\frac{20 M^2 r}{3 h^7}+\Order{r}^2.$$

$\bullet$ For other values of $n$ one can calculate the integral (\ref{msol}) numerically. Then, studying various numerical solutions at different values of the parameters, we observe that the black hole solutions exist only for sufficiently dense profiles, corresponding to small values of $h$. The maximum value of $h/M$, for which the event horizon exists, decreases with $n$.

For small $n$, the halo density decreases fast and the black-hole metric outside the horizon deviates from the Schwarzschild geometry only slightly, unless $h$ is sufficiently close to its extreme value (see Fig.~\ref{fig:smallnmetric}). Similar black-hole geometries have been recently obtained by introducing Gaussian nonlocality in the radial coordinate of the Schwarzschild metric \cite{Boos:2021kqe}.

\begin{figure*}
\resizebox{\linewidth}{!}{\includegraphics{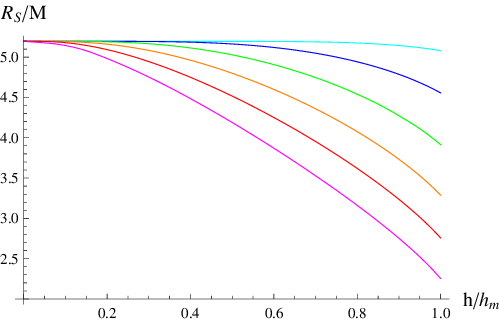}\includegraphics{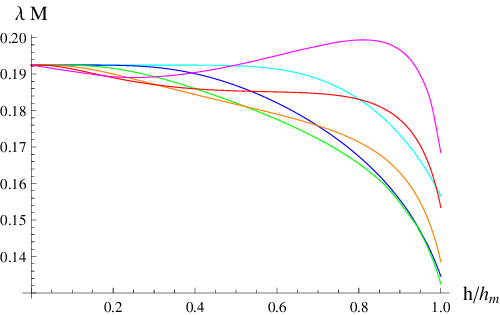}}
\caption{Shadow radius (left panel) and Lyapunov exponents (right panel) as functions of the scale parameter $h$ in units of its maximum value $h_m$ for the black hole solutions with Einasto profile (from top to bottom): $n=1$ (cyan), $n=2$ (blue), $n=3$ (green), $n=4$ (orange), $n=5$ (red), $n=6$ (magenta). The value $h=0$ corresponds to the Schwarzschild black hole.}\label{fig:Einastoshadow}
\end{figure*}

For large $n$ the Einasto density profile must be very dense and decays relatively slowly. Therefore, the black-hole metric is significantly different from the Schwarzschild geometry within the parametric range of $h$ providing the existence of the event horizon (see Fig.~\ref{fig:largenmetric}). As a result for larger $n$ the shadow radius and Lyapunov exponents generally differ more from their Schwarzschild values (see Fig.~\ref{fig:Einastoshadow}), which correspond to the limit $h\to0$.

\section{Dehnen-type distributions}\label{sec:Dehnen}
Another family of density profiles, which generalizes several well-known halo profiles through tunable slope parameters, is defined \cite{Dehnen:1993uh,Taylor:2002zd}
\begin{equation}\label{Dehnendensity}
\rho(r)=\rho_0 \left(\frac{r}{a}\right)^{-\alpha} \left(1+\frac{r^k}{a^k}\right)^{-(\gamma-\alpha)/k}.
\end{equation}
The density becomes infinite at $r=0$, unless $\alpha\leq0$. This requirement rules out a number of widely used phenomenological halo profiles, including the NFW and Hernquist models, which correspond to $\alpha=1$, as well as the Moore profile with $\alpha=7/5$ \cite{Moore:1997sg}. However, those profiles are intended to describe the dark-matter distribution only on galactic scales. The formal divergence at the center indicates that they cannot be extrapolated down to small radii: in any physically realistic setting, the density must be regularized or modified near the origin.

We shall also consider $\gamma>3$, which satisfies the condition (\ref{convergencecond}).
Then, in order to have a monotonously decreasing profile (\ref{decreasingcond}), which satisfies the weak energy conditions, $\alpha=0$. It is interesting to note that, in this case, the fluid is barotropic \cite{Sajadi:2025prp}. The total mass is
\begin{equation}\label{totalDehnenmass}
M=\frac{4\pi\rho_0a^3\Gamma\left(1+3/k\right)\Gamma\left(\gamma/k-3/k\right)}{3\Gamma\left(\gamma/k\right)}.
\end{equation}
Again, for sufficiently small values of $a/M$, the solution possesses an event horizon. In the limit $a\to0$ while keeping the total mass $M$ in Eq.~(\ref{totalDehnenmass}) fixed, the solution reduces to the Schwarzschild geometry.

It is noteworthy that, although for $\gamma = 3$ the total mass (\ref{totalDehnenmass}) diverges, the corresponding solutions remain asymptotically flat, whereas for $\gamma = 2$ they are asymptotically characterized by a solid angle defect (see \cite{Kar:2025phe} for a review).

The obtained model describes a finite-density core that gradually steepens at large radii, ensuring a rapidly decaying outer halo. Although the density profile does not reproduce the standard models, such as Hernquist or NFW profiles, it represents a family of dark matter distributions with a flat asymptotic region and finite total mass. The analytic form of the line element provides a tractable framework for studying how dark matter affects the local geometry around the regular black hole.

In particular, for $k = 1$ one obtains a relatively simple expression,
\begin{eqnarray}\label{Dehnenmetric}
f(r)&=&1-\frac{2m(r)}{r}= 1-\frac{2M}{r}
\\\nonumber
&+&\frac{2M}{r}\left(\frac{a}{a+r}\right)^{\gamma-3}\left(1+\frac{(\gamma-1)(2a+\gamma r)r}{2(a+r)^2}\right).
\end{eqnarray}
where
\begin{equation}\label{DehnenM}
M=\frac{8\pi\rho_0a^3}{(\gamma-3)(\gamma-2)(\gamma-1)}.
\end{equation}

The Ricci scalar and the Kretschmann scalar for the above metric read
\begin{eqnarray}\nonumber
&&R=\frac{M (4a-(\gamma-4)r)(\gamma-3)(\gamma-2)(\gamma-1)a^{\gamma-3}}{(a+r)^{\gamma+1}},
\\\nonumber&&
\!\!\!\!R_{\mu\nu\lambda\sigma}R^{\mu\nu\lambda\sigma}=\frac{8M^2(\gamma-3)^2(\gamma-2)^2(\gamma-1)^2}{3a^6}+\Order{r}.
\end{eqnarray}

For $\gamma=4$ the metric function (\ref{Dehnenmetric}) reads
\begin{equation}
f(r)=1-\frac{2m(r)}{r}= 1-\frac{2Mr^2}{(r+a)^3}.
\end{equation}

For $\gamma=4$ and $k=2$, we have a regular black hole solution, which can be obtained within nonlinear electrodynamics minimally coupled to gravity \cite{Dymnikova:2004zc}.

Another simple solution we obtain for $k=3$,
\begin{equation}\label{Dehnenk3metric}
f(r)=1-\frac{2m(r)}{r}= 1-\frac{2M}{r}+\frac{2M}{r}\left(1+\frac{r^3}{a^3}\right)^{1-\gamma/3},
\end{equation}
where
\begin{equation}
M=\frac{4\pi\rho_0a^3}{\gamma-3}.
\end{equation}

The curvature invariants for the above metric are
\begin{eqnarray}\nonumber
&&R=\frac{2 M (\gamma -3)\left(4 a^3-r^3 (\gamma -4)\right)}{a^6}\left(1+\frac{r^3}{a^3}\right)^{-1-\gamma /3},
\\\nonumber&&
R_{\mu\nu\lambda\sigma}R^{\mu\nu\lambda\sigma}=\frac{32M^2(\gamma-3)^2}{3a^6}+\Order{r}^3.
\end{eqnarray}
For $\gamma=4$ we obtain the solution~(3.18) of \cite{Kar:2025phe}. For $\gamma=6$ the solution (\ref{Dehnenk3metric}) coincides with the Hayward black hole \cite{Hayward:2005gi}.

\section{Axial perturbations}\label{sec:axial}
In \cite{Chakraborty:2024gcr} two types of linear axial (odd) perturbations of the spherically symmetric configurations with anisotropic fluid have been derived. In both cases the perturbation equations can be reduced to the wave-like form:
\begin{equation}
\frac{d^2 \Psi}{dr_*^2} + \left( \omega^2 - V(r) \right) \Psi = 0, \quad dr_* \equiv \frac{dr}{f(r)}.
\end{equation}
The ``up'' perturbations correspond to the the choice of the observer, which does not see a variation in the contravariant components of the fluid and the ``down'' perturbations correspond the absence of the variations in the covariant components. The corresponding effective potentials are different:
\begin{eqnarray}
V^\text{(up)}(r)&=&f(r)\Biggl(\frac{\ell(\ell+1)}{r^2}-\frac{6m(r)}{r^3}
\\\nonumber&&
+4\pi\left[\rho(r)-5P_r(r)+4P(r)\right]\Biggr),\\
V^\text{(down)}(r)&=&f(r)\Biggl(\frac{\ell(\ell+1)}{r^2}-\frac{6m(r)}{r^3}
\\\nonumber&&
+4\pi\rho(r)-4\pi P_r(r)\Biggr).
\end{eqnarray}
Taking into account the condition (\ref{vacuum}) and (\ref{Pex}) we find
\begin{eqnarray}
V^\text{(up)}(r)&=&f(r)\Biggl(\frac{\ell(\ell+1)}{r^2}-\frac{6m(r)}
{r^3}+8\pi\rho(r)
\nonumber\\&&
-8\pi r\rho'(r)\Biggr),\\
V^\text{(down)}(r)&=&f(r)\Biggl(\frac{\ell(\ell+1)}{r^2}-\frac{6m(r)}{r^3}+8\pi\rho(r)\Biggr).\nonumber
\end{eqnarray}
Since $\rho(r)>0$ and $\rho'(r)<0$, both effective potentials are positive definite outside the event horizon (as $2m(r)\leq r$) for $\ell\geq2$. Thus, the differential operator
\begin{equation}
\mathcal{D} = -\frac{\partial^2}{\partial r_*^2} + V(r)
\end{equation}
is a positive self-adjoint operator in the Hilbert space of square integrable functions of \(r_*\). Therefore, all solutions of the perturbation equations with compact support initial conditions are bounded \cite{Konoplya:2011qq}, that is, the configuration is stable against axial-type gravitational perturbations. In other words, any such perturbation of the black hole surrounded by the fluid decays and eventually settles back to the unperturbed background configuration. Hence, no growing mode must be present in the axial sector.

\section{Black–hole shadow and Lyapunov exponent}\label{sec:shadow}
One of the most important observable characteristics of a black hole surrounded by its galactic environment is the radius of its shadow.
Consequently, a substantial body of research has been devoted to studying black hole shadows in various theories of gravity and for different types of surrounding matter and environments (see ~\cite{Cunha:2018acu,Perlick:2021aok,Bambi:2015kza,Hioki:2009na,Bonanno:2025dry,Konoplya:2019sns,Younsi:2016azx,Bambi:2019tjh,Konoplya:2025bte,Perlick:2015vta,Konoplya:2024lch,Khodadi:2020jij,Kumar:2018ple,Konoplya:2021slg,Afrin:2021imp,Zakharov:2014lqa,Zakharov:2025rhb,Zakharov:2025cnq,Lutfuoglu:2025ldc,Konoplya:2019fpy,Konoplya:2019goy,Tsukamoto:2017fxq,Stashko:2024wuq,Spina:2025wxb} and literature therein for reviews and recent examples.

The timelike and axial Killing vectors yield the conserved energy and angular momentum
\begin{equation}
E \equiv f(r)\,\frac{dt}{d\lambda},\qquad
L \equiv r^{2}\,\frac{d\phi}{d\lambda},
\end{equation}
with $\lambda$ an affine parameter. Using $ds^{2}=0$ for null rays in the equatorial plane $\theta=\pi/2$,
\begin{equation}
0=-f(r)\left(\frac{dt}{d\lambda}\right)^{2}
+\frac{B^{2}(r)}{f(r)}\left(\frac{dr}{d\lambda}\right)^{2}
+r^{2}\left(\frac{d\phi}{d\lambda}\right)^{2},\label{eq:nullequatorial}
\end{equation}
one finds the radial equation
\begin{equation}
\left(\frac{dr}{d\lambda}\right)^{2}=\frac{E^{2}}{B^{2}(r)}\left[1-\frac{f(r)\,b^{2}}{r^{2}}\right],
\qquad
b\equiv \frac{L}{E}\,,
\label{eq:radial_null}
\end{equation}
where $b$ is the (asymptotic) impact parameter.

Null circular orbit satisfies the following relation
\begin{equation}
\frac{dr}{d\lambda}=0 \quad\Longrightarrow\quad b^{2}=\frac{r^{2}}{f(r)}.
\end{equation}
Extremizing the impact parameter we obtain the photon sphere radius, $r_{\rm ph}$, as the solution of the following equation
\begin{equation}
\frac{db}{dr}\Big|_{r=r_{\rm ph}}=0
\;\;\Longleftrightarrow\;\;
r_{\rm ph}\,f'(r_{\rm ph})-2\,f(r_{\rm ph})=0.
\label{eq:photon_sphere_condition}
\end{equation}
The corresponding \emph{critical impact parameter} coincides with shadow radius for an observer at infinity,
\begin{equation}
R_S=b\Big|_{r=r_{\rm ph}}
=\frac{r_{\rm ph}}{\sqrt{f(r_{\rm ph})}}.
\label{eq:bc}
\end{equation}
Notice that $B(r)$ drops out of Eqs.~(\ref{eq:photon_sphere_condition}) and~(\ref{eq:bc}); only $f(r)$ determines the photon sphere and the shadow radius $R_S$.

By substituting $r=r_{\rm ph}+\delta r$ into (\ref{eq:nullequatorial}) we obtain the equation for the radial coordinate
of the photons that are leaving the circular orbit,
\begin{equation}
\left(\frac{d}{dt}\delta r\right)^{2}=\lambda^2 \delta r^2+\Order{\delta r}^3,
\end{equation}
where $\lambda$ is the the Lyapunov exponent, satisfying,
\begin{equation}
\lambda^2=-\frac{r^2f(r)}{2B^2(r)}\frac{d^2}{dr^2}\frac{f(r)}{r^2}\Big|_{r=r_{\rm ph}}.
\end{equation}
Note that $B(r)=1$ for all the configurations considered in the present paper because of the condition (\ref{Bsol}).

For large $n$ the Einasto density profile must be very dense and decays relatively slowly. Therefore, the black-hole metric is significantly different from the Schwarzschild geometry within the parametric range of $h$ providing the existence of the event horizon (see Fig.~\ref{fig:largenmetric}). As a result for larger $n$ the shadow radius and Lyapunov exponents generally differ more from their Schwarzschild values (see Fig.~\ref{fig:Einastoshadow}).

\begin{figure*}
\resizebox{\linewidth}{!}{\includegraphics{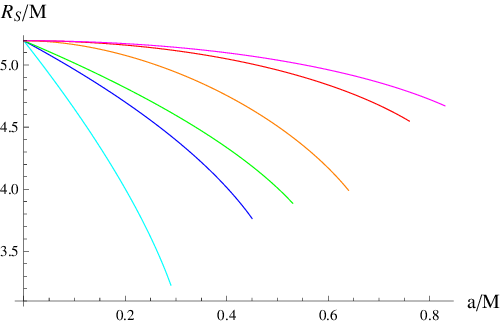}\includegraphics{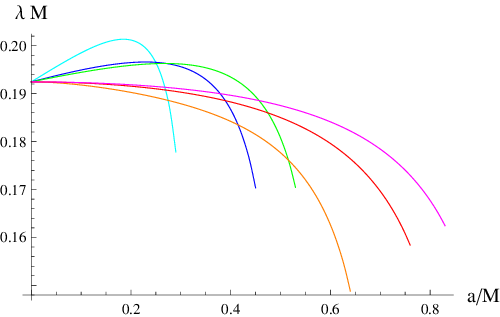}}
\caption{Shadow radius (left panel) and Lyapunov exponents (right panel) as functions of the scale parameter $a$ for the black hole solutions with Dehnen-type profiles for $\alpha=0$ (from shorter to longer region of $a$): $\gamma=4$, $k=1$ (cyan), $\gamma=4$, $k=2$ (blue), $\gamma=4$, $k=3$ (green), $\gamma=5$, $k=1$ (orange), $\gamma=5$, $k=2$ (red), $\gamma=5$, $k=3$ (magenta). The value $a=0$ corresponds to the Schwarzschild black hole.}\label{fig:Dehnenshadow}
\end{figure*}

For Dehnen-type profiles with $\alpha=0$ (see Fig.~\ref{fig:Dehnenshadow}), the halo density is higher for smaller values of $\gamma$ and also slightly increases as the parameter $k$ decreases. Consequently, the range of black hole shadow radii becomes larger for smaller $\gamma$ and $k$, whereas the presence of the event horizon is observed as the scale parameter $a$ varies within a comparatively narrower interval. In contrast, the Lyapunov exponent exhibits a broader variation range for higher values of $\gamma$, indicating stronger sensitivity of the photon orbit stability to changes in the halo profile.

It should be noted that the computed shadow radii and Lyapunov exponents can also be used to estimate quasinormal modes in the eikonal (high-multipole) limit through the well-known correspondence between the orbital frequency/shadow's radius of the circular null geodesic and the real part of the quasinormal frequency, as well as between the Lyapunov exponent and the damping rate~\cite{Cardoso:2008bp,Jusufi:2020dhz}. Although several exceptions to this correspondence have been identified in~\cite{Konoplya:2017wot,Khanna:2016yow,Konoplya:2022gjp,Bolokhov:2023dxq,Konoplya:2020bxa,Konoplya:2025afm}, the present wave equation exhibits proper WKB eikonal behavior, ensuring that the correspondence remains valid in our case.

\section{Discussions}\label{sec:discussions}
It is important to emphasize that, once we assume that $m(0)=0$ in (\ref{msol}), the considered solutions always have a de Sitter core,
\begin{eqnarray}
f(r)=1-\frac{2m(r)}{r}&=&1-\frac{8\pi\rho_0}{3} r^2+\Order{r}^3.
\end{eqnarray}
Therefore, the Ricci scalar and Kretschmann scalar,
\begin{eqnarray}
R&=&-f''(r)-\frac{4}{r}f'(r)+\frac{2\left(1-f(r)\right)}{r^2},
\\\nonumber
R_{\mu\nu\lambda\sigma}R^{\mu\nu\lambda\sigma}&=&f''(r)^2+\frac{4}{r^2}f'(r)^2+\frac{4}{r^4}\left(1-f(r)\right)^2,
\end{eqnarray}
are regular at $r=0$.

The general static spherically symmetric solution of the Einstein equations discussed in Sec.~\ref{sec:basic}can be parametrized by the arbitrary constant $m(0)=m_0$. The resulting metric function $\widetilde{f}(r)$ can be expressed as follows:
\begin{equation}
\widetilde{f}(r)=f(r)-\frac{2m_0}{r},
\end{equation}
leading to a singular family of black holes with only one regular solution corresponding to $m_0=0$. A similar observation has been made for the Bardeen and Hayward black holes as solutions within nonlinear electrodynamics \cite{Huang:2025uhv}. Energy conditions of singular black holes immersed in various models of dark matter halo, including the Einasto and Dehnan density profiles, were analyzed in \cite{Datta:2023zmd}.

In order to regard regular black holes supported by a dark-matter halo as astrophysically plausible, one must assume that such halo configurations could have existed already during the epoch of stellar collapse. In standard scenarios of stellar evolution, the baryonic density of the collapsing star vastly exceeds the local density of dark matter, so the latter is not expected to dominate the collapse dynamics. Nevertheless, cosmological N-body simulations in the $\Lambda$CDM paradigm predict that galaxies and their progenitors formed within extended dark matter halos \cite{Navarro:1996gj,Springel:2005nw}, implying that any stellar collapse took place in an environment permeated by dark matter.
In this context, dark matter may become relevant for constructing regular black-hole interiors. Moreover, in the early Universe dark matter over densities or self-interacting dark matter models may have led to locally enhanced densities \cite{Spergel:1999mh,Tulin:2017ara}, while compact dark matter structures such as solitonic cores or spikes around primordial black holes could also provide significant local contributions \cite{Gondolo:1999ef,Ullio:2001fb}. Consequently, the dark-matter distribution could be incorporated into the stress-energy tensor sourcing the black-hole metric rather than treated as a negligible external background. Regular primordial black holes, in their turn, can act as cosmic expansion accelerators \cite{Dialektopoulos:2025mfz}. Even if dark matter did not directly drive the collapse, subsequent accretion from the ambient halo can build up dense distributions around newly formed black holes \cite{Bertone:2005xz}. Thus, although baryonic matter is the primary agent in stellar collapse, there exist several physically motivated scenarios in which dark matter halos may consistently act as effective sources in constructing regular black-hole geometries.

While the above particular models of dark matter distributions might be questioned at the stage of black-hole formation, we do not rely on these specific choices. In fact, the emergence of a regular black-hole geometry follows from rather generic assumptions about the effective matter distribution, as outlined in Sec.~\ref{sec:basic}, namely, that the density remains finite and regular everywhere, and that the anisotropic fluid satisfies the weak energy condition. Unlike the vacuum case, these minimal conditions admit regular black-hole solutions for sufficiently dense matter configurations.

\section{Conclusions}\label{sec:conclusions}
In this work we have constructed exact black hole solutions free from curvature singularities, sourced by a galactic dark-matter halo. The regularity of these geometries follows from the specific relation $P_{r}=-\rho$ between radial pressure and density, which is consistent with the phenomenological freedom in halo modeling. The regular black hole solution is shown to be stable against axial perturbations. While this could be an indication that the solution is stable, the full analysis of stability must also include the polar type of perturbations. The obtained solutions demonstrate that halo distributions can serve not only as astrophysical environments but also as effective sources of singularity resolution. The radius of shadow and the Lyapunov exponent for circular null geodesics are discussed for particular examples of regular black holes.

A natural direction for further study is the investigation of physical observables associated with these regular configurations \cite{Vagnozzi:2022moj}. In particular, the analysis of gravitational spectra through quasinormal modes and echoes can provide insights into the stability and response of such black holes under perturbations. Similarly, the study of electromagnetic spectra, including gravitational lensing \cite{Boos:2025nzc}, and energy emission, offers potential observational signatures that could distinguish regular black holes embedded in halos from their singular Schwarzschild or Kerr counterparts.

The literature on singular black holes surrounded by galactic halos is already extensive, covering quasinormal ringing \cite{Konoplya:2021ube,Konoplya:2022hbl,Dubinsky:2025fwv,Feng:2025iao,Pezzella:2024tkf,Chakraborty:2024gcr,Liu:2024bfj,Liu:2024xcd,Zhao:2023tyo,Daghigh:2022pcr,Zhang:2021bdr}, grey-body factors \cite{Hamil:2025pte,Mollicone:2024lxy,Tovar:2025apz,Lutfuoglu:2025kqp,Pathrikar:2025sin}, black hole shadows and lensing \cite{Konoplya:2025nqv,Hou:2018avu,Kouniatalis:2025itj,Fernandes:2025osu,Chen:2024lpd,Tan:2024hzw,Macedo:2024qky,Xavier:2023exm,Figueiredo:2023gas,Konoplya:2025mvj}, and dynamical processes such as accretion \cite{Chowdhury:2025tpt,Heydari-Fard:2024wgu}. Extending these analyses to the case of regular black holes supported by halo matter would allow one to quantify the extent to which regularity alters these spectra and potentially leaves imprints observable with current or near-future instruments.

\acknowledgments
A.~Z.~was supported by Conselho Nacional de Desenvolvimento Científico e Tecnológico (CNPq).

\bibliography{BHEinasto}

\begin{thebibliography}{152}%
\makeatletter
\providecommand \@ifxundefined [1]{%
 \@ifx{#1\undefined}
}%
\providecommand \@ifnum [1]{%
 \ifnum #1\expandafter \@firstoftwo
 \else \expandafter \@secondoftwo
 \fi
}%
\providecommand \@ifx [1]{%
 \ifx #1\expandafter \@firstoftwo
 \else \expandafter \@secondoftwo
 \fi
}%
\providecommand \natexlab [1]{#1}%
\providecommand \enquote  [1]{``#1''}%
\providecommand \bibnamefont  [1]{#1}%
\providecommand \bibfnamefont [1]{#1}%
\providecommand \citenamefont [1]{#1}%
\providecommand \href@noop [0]{\@secondoftwo}%
\providecommand \href [0]{\begingroup \@sanitize@url \@href}%
\providecommand \@href[1]{\@@startlink{#1}\@@href}%
\providecommand \@@href[1]{\endgroup#1\@@endlink}%
\providecommand \@sanitize@url [0]{\catcode `\\12\catcode `\$12\catcode
  `\&12\catcode `\#12\catcode `\^12\catcode `\_12\catcode `\%12\relax}%
\providecommand \@@startlink[1]{}%
\providecommand \@@endlink[0]{}%
\providecommand \url  [0]{\begingroup\@sanitize@url \@url }%
\providecommand \@url [1]{\endgroup\@href {#1}{\urlprefix }}%
\providecommand \urlprefix  [0]{URL }%
\providecommand \Eprint [0]{\href }%
\providecommand \doibase [0]{http://dx.doi.org/}%
\providecommand \selectlanguage [0]{\@gobble}%
\providecommand \bibinfo  [0]{\@secondoftwo}%
\providecommand \bibfield  [0]{\@secondoftwo}%
\providecommand \translation [1]{[#1]}%
\providecommand \BibitemOpen [0]{}%
\providecommand \bibitemStop [0]{}%
\providecommand \bibitemNoStop [0]{.\EOS\space}%
\providecommand \EOS [0]{\spacefactor3000\relax}%
\providecommand \BibitemShut  [1]{\csname bibitem#1\endcsname}%
\let\auto@bib@innerbib\@empty
\bibitem [{\citenamefont {Bardeen}(1968)}]{Bardeen:1968}%
  \BibitemOpen
  \bibfield  {author} {\bibinfo {author} {\bibfnamefont {J.~M.}\ \bibnamefont
  {Bardeen}},\ }in\ \href@noop {} {\emph {\bibinfo {booktitle} {{Proceedings of
  the International Conference GR5}}}}\ (\bibinfo {address} {Tbilisi, USSR},\
  \bibinfo {year} {1968})\BibitemShut {NoStop}%
\bibitem [{\citenamefont {Hayward}(2006)}]{Hayward:2005gi}%
  \BibitemOpen
  \bibfield  {author} {\bibinfo {author} {\bibfnamefont {S.~A.}\ \bibnamefont
  {Hayward}},\ }\href {\doibase 10.1103/PhysRevLett.96.031103} {\bibfield
  {journal} {\bibinfo  {journal} {Phys. Rev. Lett.}\ }\textbf {\bibinfo
  {volume} {96}},\ \bibinfo {pages} {031103} (\bibinfo {year} {2006})},\
  \Eprint {http://arxiv.org/abs/gr-qc/0506126} {arXiv:gr-qc/0506126}
  \BibitemShut {NoStop}%
\bibitem [{\citenamefont {Simpson}\ and\ \citenamefont
  {Visser}(2019)}]{Simpson:2018tsi}%
  \BibitemOpen
  \bibfield  {author} {\bibinfo {author} {\bibfnamefont {A.}~\bibnamefont
  {Simpson}}\ and\ \bibinfo {author} {\bibfnamefont {M.}~\bibnamefont
  {Visser}},\ }\href {\doibase 10.1088/1475-7516/2019/02/042} {\bibfield
  {journal} {\bibinfo  {journal} {JCAP}\ }\textbf {\bibinfo {volume} {02}},\
  \bibinfo {pages} {042} (\bibinfo {year} {2019})},\ \Eprint
  {http://arxiv.org/abs/1812.07114} {arXiv:1812.07114 [gr-qc]} \BibitemShut
  {NoStop}%
\bibitem [{\citenamefont {Ansoldi}(2008)}]{Ansoldi:2008jw}%
  \BibitemOpen
  \bibfield  {author} {\bibinfo {author} {\bibfnamefont {S.}~\bibnamefont
  {Ansoldi}},\ }in\ \href@noop {} {\emph {\bibinfo {booktitle} {{Conference on
  Black Holes and Naked Singularities}}}}\ (\bibinfo {year} {2008})\ \Eprint
  {http://arxiv.org/abs/0802.0330} {arXiv:0802.0330 [gr-qc]} \BibitemShut
  {NoStop}%
\bibitem [{\citenamefont {Ayon-Beato}\ and\ \citenamefont
  {Garcia}(1998)}]{AyonBeato:1998ub}%
  \BibitemOpen
  \bibfield  {author} {\bibinfo {author} {\bibfnamefont {E.}~\bibnamefont
  {Ayon-Beato}}\ and\ \bibinfo {author} {\bibfnamefont {A.}~\bibnamefont
  {Garcia}},\ }\href {\doibase 10.1103/PhysRevLett.80.5056} {\bibfield
  {journal} {\bibinfo  {journal} {Phys. Rev. Lett.}\ }\textbf {\bibinfo
  {volume} {80}},\ \bibinfo {pages} {5056} (\bibinfo {year} {1998})},\ \Eprint
  {http://arxiv.org/abs/gr-qc/9911046} {arXiv:gr-qc/9911046} \BibitemShut
  {NoStop}%
\bibitem [{\citenamefont {Dymnikova}(1992)}]{Dymnikova:1992ux}%
  \BibitemOpen
  \bibfield  {author} {\bibinfo {author} {\bibfnamefont {I.}~\bibnamefont
  {Dymnikova}},\ }\href@noop {} {\bibfield  {journal} {\bibinfo  {journal}
  {Gen. Rel. Grav.}\ }\textbf {\bibinfo {volume} {24}},\ \bibinfo {pages} {235}
  (\bibinfo {year} {1992})}\BibitemShut {NoStop}%
\bibitem [{\citenamefont {Bronnikov}(2001)}]{Bronnikov:2000vy}%
  \BibitemOpen
  \bibfield  {author} {\bibinfo {author} {\bibfnamefont {K.~A.}\ \bibnamefont
  {Bronnikov}},\ }\href {\doibase 10.1103/PhysRevD.63.044005} {\bibfield
  {journal} {\bibinfo  {journal} {Phys. Rev. D}\ }\textbf {\bibinfo {volume}
  {63}},\ \bibinfo {pages} {044005} (\bibinfo {year} {2001})},\ \Eprint
  {http://arxiv.org/abs/gr-qc/0006014} {arXiv:gr-qc/0006014} \BibitemShut
  {NoStop}%
\bibitem [{\citenamefont {Bronnikov}(2024)}]{Bronnikov:2024izh}%
  \BibitemOpen
  \bibfield  {author} {\bibinfo {author} {\bibfnamefont {K.~A.}\ \bibnamefont
  {Bronnikov}},\ }\href {\doibase 10.1103/PhysRevD.110.024021} {\bibfield
  {journal} {\bibinfo  {journal} {Phys. Rev. D}\ }\textbf {\bibinfo {volume}
  {110}},\ \bibinfo {pages} {024021} (\bibinfo {year} {2024})},\ \Eprint
  {http://arxiv.org/abs/2404.14816} {arXiv:2404.14816 [gr-qc]} \BibitemShut
  {NoStop}%
\bibitem [{\citenamefont {Bronnikov}\ and\ \citenamefont
  {Fabris}(2006)}]{Bronnikov:2005gm}%
  \BibitemOpen
  \bibfield  {author} {\bibinfo {author} {\bibfnamefont {K.~A.}\ \bibnamefont
  {Bronnikov}}\ and\ \bibinfo {author} {\bibfnamefont {J.~C.}\ \bibnamefont
  {Fabris}},\ }\href {\doibase 10.1103/PhysRevLett.96.251101} {\bibfield
  {journal} {\bibinfo  {journal} {Phys. Rev. Lett.}\ }\textbf {\bibinfo
  {volume} {96}},\ \bibinfo {pages} {251101} (\bibinfo {year} {2006})},\
  \Eprint {http://arxiv.org/abs/gr-qc/0511109} {arXiv:gr-qc/0511109}
  \BibitemShut {NoStop}%
\bibitem [{\citenamefont {Kazakov}\ and\ \citenamefont
  {Solodukhin}(1994)}]{Kazakov:1993ha}%
  \BibitemOpen
  \bibfield  {author} {\bibinfo {author} {\bibfnamefont {D.~I.}\ \bibnamefont
  {Kazakov}}\ and\ \bibinfo {author} {\bibfnamefont {S.~N.}\ \bibnamefont
  {Solodukhin}},\ }\href {\doibase 10.1016/S0550-3213(94)80045-6} {\bibfield
  {journal} {\bibinfo  {journal} {Nucl. Phys. B}\ }\textbf {\bibinfo {volume}
  {429}},\ \bibinfo {pages} {153} (\bibinfo {year} {1994})},\ \Eprint
  {http://arxiv.org/abs/hep-th/9310150} {arXiv:hep-th/9310150} \BibitemShut
  {NoStop}%
\bibitem [{\citenamefont {Modesto}(2009)}]{Modesto:2008jz}%
  \BibitemOpen
  \bibfield  {author} {\bibinfo {author} {\bibfnamefont {L.}~\bibnamefont
  {Modesto}},\ }\href {\doibase 10.1088/0264-9381/26/24/242002} {\bibfield
  {journal} {\bibinfo  {journal} {Class. Quant. Grav.}\ }\textbf {\bibinfo
  {volume} {26}},\ \bibinfo {pages} {242002} (\bibinfo {year} {2009})},\
  \Eprint {http://arxiv.org/abs/0812.2214} {arXiv:0812.2214 [gr-qc]}
  \BibitemShut {NoStop}%
\bibitem [{\citenamefont {Bonanno}\ and\ \citenamefont
  {Reuter}(2000)}]{Bonanno:2000ep}%
  \BibitemOpen
  \bibfield  {author} {\bibinfo {author} {\bibfnamefont {A.}~\bibnamefont
  {Bonanno}}\ and\ \bibinfo {author} {\bibfnamefont {M.}~\bibnamefont
  {Reuter}},\ }\href {\doibase 10.1103/PhysRevD.62.043008} {\bibfield
  {journal} {\bibinfo  {journal} {Phys. Rev. D}\ }\textbf {\bibinfo {volume}
  {62}},\ \bibinfo {pages} {043008} (\bibinfo {year} {2000})},\ \Eprint
  {http://arxiv.org/abs/hep-th/0002196} {arXiv:hep-th/0002196} \BibitemShut
  {NoStop}%
\bibitem [{\citenamefont {Lan}\ \emph {et~al.}(2023)\citenamefont {Lan},
  \citenamefont {Yang}, \citenamefont {Guo},\ and\ \citenamefont
  {Miao}}]{Lan:2023cvz}%
  \BibitemOpen
  \bibfield  {author} {\bibinfo {author} {\bibfnamefont {C.}~\bibnamefont
  {Lan}}, \bibinfo {author} {\bibfnamefont {H.}~\bibnamefont {Yang}}, \bibinfo
  {author} {\bibfnamefont {Y.}~\bibnamefont {Guo}}, \ and\ \bibinfo {author}
  {\bibfnamefont {Y.-G.}\ \bibnamefont {Miao}},\ }\href {\doibase
  10.1007/s10773-023-05454-1} {\bibfield  {journal} {\bibinfo  {journal} {Int.
  J. Theor. Phys.}\ }\textbf {\bibinfo {volume} {62}},\ \bibinfo {pages} {202}
  (\bibinfo {year} {2023})},\ \Eprint {http://arxiv.org/abs/2303.11696}
  {arXiv:2303.11696 [gr-qc]} \BibitemShut {NoStop}%
\bibitem [{\citenamefont {Bonanno}\ \emph {et~al.}(2024)\citenamefont
  {Bonanno}, \citenamefont {Malafarina},\ and\ \citenamefont
  {Panassiti}}]{Bonanno:2023rzk}%
  \BibitemOpen
  \bibfield  {author} {\bibinfo {author} {\bibfnamefont {A.}~\bibnamefont
  {Bonanno}}, \bibinfo {author} {\bibfnamefont {D.}~\bibnamefont {Malafarina}},
  \ and\ \bibinfo {author} {\bibfnamefont {A.}~\bibnamefont {Panassiti}},\
  }\href {\doibase 10.1103/PhysRevLett.132.031401} {\bibfield  {journal}
  {\bibinfo  {journal} {Phys. Rev. Lett.}\ }\textbf {\bibinfo {volume} {132}},\
  \bibinfo {pages} {031401} (\bibinfo {year} {2024})},\ \Eprint
  {http://arxiv.org/abs/2308.10890} {arXiv:2308.10890 [gr-qc]} \BibitemShut
  {NoStop}%
\bibitem [{\citenamefont {Bonanno}\ \emph {et~al.}(2025)\citenamefont
  {Bonanno}, \citenamefont {Konoplya}, \citenamefont {Oglialoro},\ and\
  \citenamefont {Spina}}]{Bonanno:2025dry}%
  \BibitemOpen
  \bibfield  {author} {\bibinfo {author} {\bibfnamefont {A.~M.}\ \bibnamefont
  {Bonanno}}, \bibinfo {author} {\bibfnamefont {R.~A.}\ \bibnamefont
  {Konoplya}}, \bibinfo {author} {\bibfnamefont {G.}~\bibnamefont {Oglialoro}},
  \ and\ \bibinfo {author} {\bibfnamefont {A.}~\bibnamefont {Spina}},\ }\href
  {\doibase 10.1088/1475-7516/2025/12/042} {\bibfield  {journal} {\bibinfo
  {journal} {JCAP}\ }\textbf {\bibinfo {volume} {12}},\ \bibinfo {pages} {042}
  (\bibinfo {year} {2025})},\ \Eprint {http://arxiv.org/abs/2509.12469}
  {arXiv:2509.12469 [gr-qc]} \BibitemShut {NoStop}%
\bibitem [{\citenamefont {Spina}(2025)}]{Spina:2025wxb}%
  \BibitemOpen
  \bibfield  {author} {\bibinfo {author} {\bibfnamefont {A.}~\bibnamefont
  {Spina}},\ }\href {\doibase 10.53941/ijgtp.2025.100008} {\bibfield  {journal}
  {\bibinfo  {journal} {Int. J. Grav. Theor. Phys.}\ }\textbf {\bibinfo
  {volume} {1}},\ \bibinfo {pages} {8} (\bibinfo {year} {2025})},\ \Eprint
  {http://arxiv.org/abs/2510.14552} {arXiv:2510.14552 [gr-qc]} \BibitemShut
  {NoStop}%
\bibitem [{\citenamefont {Zhang}\ \emph {et~al.}(2025)\citenamefont {Zhang},
  \citenamefont {Lewandowski}, \citenamefont {Ma},\ and\ \citenamefont
  {Yang}}]{Zhang:2024ney}%
  \BibitemOpen
  \bibfield  {author} {\bibinfo {author} {\bibfnamefont {C.}~\bibnamefont
  {Zhang}}, \bibinfo {author} {\bibfnamefont {J.}~\bibnamefont {Lewandowski}},
  \bibinfo {author} {\bibfnamefont {Y.}~\bibnamefont {Ma}}, \ and\ \bibinfo
  {author} {\bibfnamefont {J.}~\bibnamefont {Yang}},\ }\href {\doibase
  10.1103/d6ks-d576} {\bibfield  {journal} {\bibinfo  {journal} {Phys. Rev. D}\
  }\textbf {\bibinfo {volume} {112}},\ \bibinfo {pages} {044054} (\bibinfo
  {year} {2025})},\ \Eprint {http://arxiv.org/abs/2412.02487} {arXiv:2412.02487
  [gr-qc]} \BibitemShut {NoStop}%
\bibitem [{\citenamefont {Solodukhin}\ and\ \citenamefont
  {Tagiev}(2025)}]{Solodukhin:2025opw}%
  \BibitemOpen
  \bibfield  {author} {\bibinfo {author} {\bibfnamefont {S.~N.}\ \bibnamefont
  {Solodukhin}}\ and\ \bibinfo {author} {\bibfnamefont {V.}~\bibnamefont
  {Tagiev}},\ }\href@noop {} {\  (\bibinfo {year} {2025})},\ \Eprint
  {http://arxiv.org/abs/2511.03879} {arXiv:2511.03879 [gr-qc]} \BibitemShut
  {NoStop}%
\bibitem [{\citenamefont {Bueno}\ \emph
  {et~al.}(2025{\natexlab{a}})\citenamefont {Bueno}, \citenamefont {Cano},\
  and\ \citenamefont {Hennigar}}]{Bueno:2024dgm}%
  \BibitemOpen
  \bibfield  {author} {\bibinfo {author} {\bibfnamefont {P.}~\bibnamefont
  {Bueno}}, \bibinfo {author} {\bibfnamefont {P.~A.}\ \bibnamefont {Cano}}, \
  and\ \bibinfo {author} {\bibfnamefont {R.~A.}\ \bibnamefont {Hennigar}},\
  }\href {\doibase 10.1016/j.physletb.2025.139260} {\bibfield  {journal}
  {\bibinfo  {journal} {Phys. Lett. B}\ }\textbf {\bibinfo {volume} {861}},\
  \bibinfo {pages} {139260} (\bibinfo {year} {2025}{\natexlab{a}})},\ \Eprint
  {http://arxiv.org/abs/2403.04827} {arXiv:2403.04827 [gr-qc]} \BibitemShut
  {NoStop}%
\bibitem [{\citenamefont {Bueno}\ \emph
  {et~al.}(2025{\natexlab{b}})\citenamefont {Bueno}, \citenamefont {Hennigar},
  \citenamefont {Murcia},\ and\ \citenamefont {Vicente-Cano}}]{Bueno:2025tli}%
  \BibitemOpen
  \bibfield  {author} {\bibinfo {author} {\bibfnamefont {P.}~\bibnamefont
  {Bueno}}, \bibinfo {author} {\bibfnamefont {R.~A.}\ \bibnamefont {Hennigar}},
  \bibinfo {author} {\bibfnamefont {{\'A}.~J.}\ \bibnamefont {Murcia}}, \ and\
  \bibinfo {author} {\bibfnamefont {A.}~\bibnamefont {Vicente-Cano}},\
  }\href@noop {} {\  (\bibinfo {year} {2025}{\natexlab{b}})},\ \Eprint
  {http://arxiv.org/abs/2512.19796} {arXiv:2512.19796 [gr-qc]} \BibitemShut
  {NoStop}%
\bibitem [{\citenamefont {Bueno}\ \emph
  {et~al.}(2025{\natexlab{c}})\citenamefont {Bueno}, \citenamefont {Cano},
  \citenamefont {Hennigar},\ and\ \citenamefont {Murcia}}]{Bueno:2024eig}%
  \BibitemOpen
  \bibfield  {author} {\bibinfo {author} {\bibfnamefont {P.}~\bibnamefont
  {Bueno}}, \bibinfo {author} {\bibfnamefont {P.~A.}\ \bibnamefont {Cano}},
  \bibinfo {author} {\bibfnamefont {R.~A.}\ \bibnamefont {Hennigar}}, \ and\
  \bibinfo {author} {\bibfnamefont {{\'A}.~J.}\ \bibnamefont {Murcia}},\ }\href
  {\doibase 10.1103/PhysRevLett.134.181401} {\bibfield  {journal} {\bibinfo
  {journal} {Phys. Rev. Lett.}\ }\textbf {\bibinfo {volume} {134}},\ \bibinfo
  {pages} {181401} (\bibinfo {year} {2025}{\natexlab{c}})},\ \Eprint
  {http://arxiv.org/abs/2412.02742} {arXiv:2412.02742 [gr-qc]} \BibitemShut
  {NoStop}%
\bibitem [{\citenamefont {Frolov}\ and\ \citenamefont
  {Zelnikov}(2026)}]{Frolov:2026rcm}%
  \BibitemOpen
  \bibfield  {author} {\bibinfo {author} {\bibfnamefont {V.~P.}\ \bibnamefont
  {Frolov}}\ and\ \bibinfo {author} {\bibfnamefont {A.}~\bibnamefont
  {Zelnikov}},\ }\href@noop {} {\  (\bibinfo {year} {2026})},\ \Eprint
  {http://arxiv.org/abs/2601.01861} {arXiv:2601.01861 [gr-qc]} \BibitemShut
  {NoStop}%
\bibitem [{\citenamefont {Bronnikov}\ \emph {et~al.}(2007)\citenamefont
  {Bronnikov}, \citenamefont {Melnikov},\ and\ \citenamefont
  {Dehnen}}]{Bronnikov:2006fu}%
  \BibitemOpen
  \bibfield  {author} {\bibinfo {author} {\bibfnamefont {K.~A.}\ \bibnamefont
  {Bronnikov}}, \bibinfo {author} {\bibfnamefont {V.~N.}\ \bibnamefont
  {Melnikov}}, \ and\ \bibinfo {author} {\bibfnamefont {H.}~\bibnamefont
  {Dehnen}},\ }\href {\doibase 10.1007/s10714-007-0430-6} {\bibfield  {journal}
  {\bibinfo  {journal} {Gen. Rel. Grav.}\ }\textbf {\bibinfo {volume} {39}},\
  \bibinfo {pages} {973} (\bibinfo {year} {2007})},\ \Eprint
  {http://arxiv.org/abs/gr-qc/0611022} {arXiv:gr-qc/0611022} \BibitemShut
  {NoStop}%
\bibitem [{\citenamefont {Bronnikov}\ \emph {et~al.}(2003)\citenamefont
  {Bronnikov}, \citenamefont {Melnikov},\ and\ \citenamefont
  {Dehnen}}]{Bronnikov:2003gx}%
  \BibitemOpen
  \bibfield  {author} {\bibinfo {author} {\bibfnamefont {K.~A.}\ \bibnamefont
  {Bronnikov}}, \bibinfo {author} {\bibfnamefont {V.~N.}\ \bibnamefont
  {Melnikov}}, \ and\ \bibinfo {author} {\bibfnamefont {H.}~\bibnamefont
  {Dehnen}},\ }\href {\doibase 10.1103/PhysRevD.68.024025} {\bibfield
  {journal} {\bibinfo  {journal} {Phys. Rev. D}\ }\textbf {\bibinfo {volume}
  {68}},\ \bibinfo {pages} {024025} (\bibinfo {year} {2003})},\ \Eprint
  {http://arxiv.org/abs/gr-qc/0304068} {arXiv:gr-qc/0304068} \BibitemShut
  {NoStop}%
\bibitem [{\citenamefont {Casadio}\ \emph {et~al.}(2002)\citenamefont
  {Casadio}, \citenamefont {Fabbri},\ and\ \citenamefont
  {Mazzacurati}}]{Casadio:2001jg}%
  \BibitemOpen
  \bibfield  {author} {\bibinfo {author} {\bibfnamefont {R.}~\bibnamefont
  {Casadio}}, \bibinfo {author} {\bibfnamefont {A.}~\bibnamefont {Fabbri}}, \
  and\ \bibinfo {author} {\bibfnamefont {L.}~\bibnamefont {Mazzacurati}},\
  }\href {\doibase 10.1103/PhysRevD.65.084040} {\bibfield  {journal} {\bibinfo
  {journal} {Phys. Rev. D}\ }\textbf {\bibinfo {volume} {65}},\ \bibinfo
  {pages} {084040} (\bibinfo {year} {2002})},\ \Eprint
  {http://arxiv.org/abs/gr-qc/0111072} {arXiv:gr-qc/0111072} \BibitemShut
  {NoStop}%
\bibitem [{\citenamefont {Bronnikov}\ \emph {et~al.}(2012)\citenamefont
  {Bronnikov}, \citenamefont {Konoplya},\ and\ \citenamefont
  {Zhidenko}}]{Bronnikov:2012ch}%
  \BibitemOpen
  \bibfield  {author} {\bibinfo {author} {\bibfnamefont {K.~A.}\ \bibnamefont
  {Bronnikov}}, \bibinfo {author} {\bibfnamefont {R.~A.}\ \bibnamefont
  {Konoplya}}, \ and\ \bibinfo {author} {\bibfnamefont {A.}~\bibnamefont
  {Zhidenko}},\ }\href {\doibase 10.1103/PhysRevD.86.024028} {\bibfield
  {journal} {\bibinfo  {journal} {Phys. Rev. D}\ }\textbf {\bibinfo {volume}
  {86}},\ \bibinfo {pages} {024028} (\bibinfo {year} {2012})},\ \Eprint
  {http://arxiv.org/abs/1205.2224} {arXiv:1205.2224 [gr-qc]} \BibitemShut
  {NoStop}%
\bibitem [{\citenamefont {Fernando}\ and\ \citenamefont
  {Correa}(2012)}]{Fernando:2012yw}%
  \BibitemOpen
  \bibfield  {author} {\bibinfo {author} {\bibfnamefont {S.}~\bibnamefont
  {Fernando}}\ and\ \bibinfo {author} {\bibfnamefont {J.}~\bibnamefont
  {Correa}},\ }\href {\doibase 10.1103/PhysRevD.86.064039} {\bibfield
  {journal} {\bibinfo  {journal} {Phys. Rev. D}\ }\textbf {\bibinfo {volume}
  {86}},\ \bibinfo {pages} {064039} (\bibinfo {year} {2012})},\ \Eprint
  {http://arxiv.org/abs/1208.5442} {arXiv:1208.5442 [gr-qc]} \BibitemShut
  {NoStop}%
\bibitem [{\citenamefont {Flachi}\ and\ \citenamefont
  {Lemos}(2013)}]{Flachi:2012nv}%
  \BibitemOpen
  \bibfield  {author} {\bibinfo {author} {\bibfnamefont {A.}~\bibnamefont
  {Flachi}}\ and\ \bibinfo {author} {\bibfnamefont {J.~P.~S.}\ \bibnamefont
  {Lemos}},\ }\href {\doibase 10.1103/PhysRevD.87.024034} {\bibfield  {journal}
  {\bibinfo  {journal} {Phys. Rev. D}\ }\textbf {\bibinfo {volume} {87}},\
  \bibinfo {pages} {024034} (\bibinfo {year} {2013})},\ \Eprint
  {http://arxiv.org/abs/1211.6212} {arXiv:1211.6212 [gr-qc]} \BibitemShut
  {NoStop}%
\bibitem [{\citenamefont {Li}\ \emph {et~al.}(2013)\citenamefont {Li},
  \citenamefont {Hong},\ and\ \citenamefont {Lin}}]{Li:2013fka}%
  \BibitemOpen
  \bibfield  {author} {\bibinfo {author} {\bibfnamefont {J.}~\bibnamefont
  {Li}}, \bibinfo {author} {\bibfnamefont {M.}~\bibnamefont {Hong}}, \ and\
  \bibinfo {author} {\bibfnamefont {K.}~\bibnamefont {Lin}},\ }\href {\doibase
  10.1103/PhysRevD.88.064001} {\bibfield  {journal} {\bibinfo  {journal} {Phys.
  Rev. D}\ }\textbf {\bibinfo {volume} {88}},\ \bibinfo {pages} {064001}
  (\bibinfo {year} {2013})},\ \Eprint {http://arxiv.org/abs/1308.6499}
  {arXiv:1308.6499 [gr-qc]} \BibitemShut {NoStop}%
\bibitem [{\citenamefont {Liu}\ \emph {et~al.}(2020)\citenamefont {Liu},
  \citenamefont {Zhu}, \citenamefont {Wu}, \citenamefont {Jusufi},
  \citenamefont {Jamil}, \citenamefont {Azreg-A{\"\i}nou},\ and\ \citenamefont
  {Wang}}]{Liu:2020ola}%
  \BibitemOpen
  \bibfield  {author} {\bibinfo {author} {\bibfnamefont {C.}~\bibnamefont
  {Liu}}, \bibinfo {author} {\bibfnamefont {T.}~\bibnamefont {Zhu}}, \bibinfo
  {author} {\bibfnamefont {Q.}~\bibnamefont {Wu}}, \bibinfo {author}
  {\bibfnamefont {K.}~\bibnamefont {Jusufi}}, \bibinfo {author} {\bibfnamefont
  {M.}~\bibnamefont {Jamil}}, \bibinfo {author} {\bibfnamefont
  {M.}~\bibnamefont {Azreg-A{\"\i}nou}}, \ and\ \bibinfo {author}
  {\bibfnamefont {A.}~\bibnamefont {Wang}},\ }\href {\doibase
  10.1103/PhysRevD.101.084001} {\bibfield  {journal} {\bibinfo  {journal}
  {Phys. Rev. D}\ }\textbf {\bibinfo {volume} {101}},\ \bibinfo {pages}
  {084001} (\bibinfo {year} {2020})},\ \bibinfo {note} {[Erratum: Phys.Rev.D
  103, 089902 (2021)]},\ \Eprint {http://arxiv.org/abs/2003.00477}
  {arXiv:2003.00477 [gr-qc]} \BibitemShut {NoStop}%
\bibitem [{\citenamefont {Jusufi}\ \emph {et~al.}(2021)\citenamefont {Jusufi},
  \citenamefont {Azreg-A{\"\i}nou}, \citenamefont {Jamil}, \citenamefont {Wei},
  \citenamefont {Wu},\ and\ \citenamefont {Wang}}]{Jusufi:2020odz}%
  \BibitemOpen
  \bibfield  {author} {\bibinfo {author} {\bibfnamefont {K.}~\bibnamefont
  {Jusufi}}, \bibinfo {author} {\bibfnamefont {M.}~\bibnamefont
  {Azreg-A{\"\i}nou}}, \bibinfo {author} {\bibfnamefont {M.}~\bibnamefont
  {Jamil}}, \bibinfo {author} {\bibfnamefont {S.-W.}\ \bibnamefont {Wei}},
  \bibinfo {author} {\bibfnamefont {Q.}~\bibnamefont {Wu}}, \ and\ \bibinfo
  {author} {\bibfnamefont {A.}~\bibnamefont {Wang}},\ }\href {\doibase
  10.1103/PhysRevD.103.024013} {\bibfield  {journal} {\bibinfo  {journal}
  {Phys. Rev. D}\ }\textbf {\bibinfo {volume} {103}},\ \bibinfo {pages}
  {024013} (\bibinfo {year} {2021})},\ \Eprint
  {http://arxiv.org/abs/2008.08450} {arXiv:2008.08450 [gr-qc]} \BibitemShut
  {NoStop}%
\bibitem [{\citenamefont {Toshmatov}\ \emph {et~al.}(2015)\citenamefont
  {Toshmatov}, \citenamefont {Abdujabbarov}, \citenamefont {Stuchl{\'\i}k},\
  and\ \citenamefont {Ahmedov}}]{Toshmatov:2015wga}%
  \BibitemOpen
  \bibfield  {author} {\bibinfo {author} {\bibfnamefont {B.}~\bibnamefont
  {Toshmatov}}, \bibinfo {author} {\bibfnamefont {A.}~\bibnamefont
  {Abdujabbarov}}, \bibinfo {author} {\bibfnamefont {Z.}~\bibnamefont
  {Stuchl{\'\i}k}}, \ and\ \bibinfo {author} {\bibfnamefont {B.}~\bibnamefont
  {Ahmedov}},\ }\href {\doibase 10.1103/PhysRevD.91.083008} {\bibfield
  {journal} {\bibinfo  {journal} {Phys. Rev. D}\ }\textbf {\bibinfo {volume}
  {91}},\ \bibinfo {pages} {083008} (\bibinfo {year} {2015})},\ \Eprint
  {http://arxiv.org/abs/1503.05737} {arXiv:1503.05737 [gr-qc]} \BibitemShut
  {NoStop}%
\bibitem [{\citenamefont {Toshmatov}\ \emph {et~al.}(2018)\citenamefont
  {Toshmatov}, \citenamefont {Stuchl{\'\i}k},\ and\ \citenamefont
  {Ahmedov}}]{Toshmatov:2018ell}%
  \BibitemOpen
  \bibfield  {author} {\bibinfo {author} {\bibfnamefont {B.}~\bibnamefont
  {Toshmatov}}, \bibinfo {author} {\bibfnamefont {Z.}~\bibnamefont
  {Stuchl{\'\i}k}}, \ and\ \bibinfo {author} {\bibfnamefont {B.}~\bibnamefont
  {Ahmedov}},\ }\href {\doibase 10.1103/PhysRevD.98.085021} {\bibfield
  {journal} {\bibinfo  {journal} {Phys. Rev. D}\ }\textbf {\bibinfo {volume}
  {98}},\ \bibinfo {pages} {085021} (\bibinfo {year} {2018})},\ \Eprint
  {http://arxiv.org/abs/1810.06383} {arXiv:1810.06383 [gr-qc]} \BibitemShut
  {NoStop}%
\bibitem [{\citenamefont {Toshmatov}\ \emph {et~al.}(2019)\citenamefont
  {Toshmatov}, \citenamefont {Stuchl{\'\i}k}, \citenamefont {Ahmedov},\ and\
  \citenamefont {Malafarina}}]{Toshmatov:2019gxg}%
  \BibitemOpen
  \bibfield  {author} {\bibinfo {author} {\bibfnamefont {B.}~\bibnamefont
  {Toshmatov}}, \bibinfo {author} {\bibfnamefont {Z.}~\bibnamefont
  {Stuchl{\'\i}k}}, \bibinfo {author} {\bibfnamefont {B.}~\bibnamefont
  {Ahmedov}}, \ and\ \bibinfo {author} {\bibfnamefont {D.}~\bibnamefont
  {Malafarina}},\ }\href {\doibase 10.1103/PhysRevD.99.064043} {\bibfield
  {journal} {\bibinfo  {journal} {Phys. Rev. D}\ }\textbf {\bibinfo {volume}
  {99}},\ \bibinfo {pages} {064043} (\bibinfo {year} {2019})},\ \Eprint
  {http://arxiv.org/abs/1903.03778} {arXiv:1903.03778 [gr-qc]} \BibitemShut
  {NoStop}%
\bibitem [{\citenamefont {Rinc{\'o}n}\ and\ \citenamefont
  {Santos}(2020)}]{Rincon:2020cos}%
  \BibitemOpen
  \bibfield  {author} {\bibinfo {author} {\bibfnamefont {{\'A}.}~\bibnamefont
  {Rinc{\'o}n}}\ and\ \bibinfo {author} {\bibfnamefont {V.}~\bibnamefont
  {Santos}},\ }\href {\doibase 10.1140/epjc/s10052-020-08445-2} {\bibfield
  {journal} {\bibinfo  {journal} {Eur. Phys. J. C}\ }\textbf {\bibinfo {volume}
  {80}},\ \bibinfo {pages} {910} (\bibinfo {year} {2020})},\ \Eprint
  {http://arxiv.org/abs/2009.04386} {arXiv:2009.04386 [gr-qc]} \BibitemShut
  {NoStop}%
\bibitem [{\citenamefont {Yang}\ \emph {et~al.}(2021)\citenamefont {Yang},
  \citenamefont {Liu}, \citenamefont {Xu}, \citenamefont {Xing}, \citenamefont
  {Wu},\ and\ \citenamefont {Long}}]{Yang:2021cvh}%
  \BibitemOpen
  \bibfield  {author} {\bibinfo {author} {\bibfnamefont {Y.}~\bibnamefont
  {Yang}}, \bibinfo {author} {\bibfnamefont {D.}~\bibnamefont {Liu}}, \bibinfo
  {author} {\bibfnamefont {Z.}~\bibnamefont {Xu}}, \bibinfo {author}
  {\bibfnamefont {Y.}~\bibnamefont {Xing}}, \bibinfo {author} {\bibfnamefont
  {S.}~\bibnamefont {Wu}}, \ and\ \bibinfo {author} {\bibfnamefont {Z.-W.}\
  \bibnamefont {Long}},\ }\href {\doibase 10.1103/PhysRevD.104.104021}
  {\bibfield  {journal} {\bibinfo  {journal} {Phys. Rev. D}\ }\textbf {\bibinfo
  {volume} {104}},\ \bibinfo {pages} {104021} (\bibinfo {year} {2021})},\
  \Eprint {http://arxiv.org/abs/2107.06554} {arXiv:2107.06554 [gr-qc]}
  \BibitemShut {NoStop}%
\bibitem [{\citenamefont {Franzin}\ \emph {et~al.}(2022)\citenamefont
  {Franzin}, \citenamefont {Liberati}, \citenamefont {Mazza}, \citenamefont
  {Dey},\ and\ \citenamefont {Chakraborty}}]{Franzin:2022iai}%
  \BibitemOpen
  \bibfield  {author} {\bibinfo {author} {\bibfnamefont {E.}~\bibnamefont
  {Franzin}}, \bibinfo {author} {\bibfnamefont {S.}~\bibnamefont {Liberati}},
  \bibinfo {author} {\bibfnamefont {J.}~\bibnamefont {Mazza}}, \bibinfo
  {author} {\bibfnamefont {R.}~\bibnamefont {Dey}}, \ and\ \bibinfo {author}
  {\bibfnamefont {S.}~\bibnamefont {Chakraborty}},\ }\href {\doibase
  10.1103/PhysRevD.105.124051} {\bibfield  {journal} {\bibinfo  {journal}
  {Phys. Rev. D}\ }\textbf {\bibinfo {volume} {105}},\ \bibinfo {pages}
  {124051} (\bibinfo {year} {2022})},\ \Eprint
  {http://arxiv.org/abs/2201.01650} {arXiv:2201.01650 [gr-qc]} \BibitemShut
  {NoStop}%
\bibitem [{\citenamefont {Konoplya}\ \emph {et~al.}(2022)\citenamefont
  {Konoplya}, \citenamefont {Zinhailo}, \citenamefont {Kunz}, \citenamefont
  {Stuchlik},\ and\ \citenamefont {Zhidenko}}]{Konoplya:2022hll}%
  \BibitemOpen
  \bibfield  {author} {\bibinfo {author} {\bibfnamefont {R.~A.}\ \bibnamefont
  {Konoplya}}, \bibinfo {author} {\bibfnamefont {A.~F.}\ \bibnamefont
  {Zinhailo}}, \bibinfo {author} {\bibfnamefont {J.}~\bibnamefont {Kunz}},
  \bibinfo {author} {\bibfnamefont {Z.}~\bibnamefont {Stuchlik}}, \ and\
  \bibinfo {author} {\bibfnamefont {A.}~\bibnamefont {Zhidenko}},\ }\href
  {\doibase 10.1088/1475-7516/2022/10/091} {\bibfield  {journal} {\bibinfo
  {journal} {JCAP}\ }\textbf {\bibinfo {volume} {10}},\ \bibinfo {pages} {091}
  (\bibinfo {year} {2022})},\ \Eprint {http://arxiv.org/abs/2206.14714}
  {arXiv:2206.14714 [gr-qc]} \BibitemShut {NoStop}%
\bibitem [{\citenamefont {Meng}\ and\ \citenamefont
  {Zhang}(2023)}]{Meng:2022oxg}%
  \BibitemOpen
  \bibfield  {author} {\bibinfo {author} {\bibfnamefont {K.}~\bibnamefont
  {Meng}}\ and\ \bibinfo {author} {\bibfnamefont {S.-J.}\ \bibnamefont
  {Zhang}},\ }\href {\doibase 10.1088/1361-6382/acf3c6} {\bibfield  {journal}
  {\bibinfo  {journal} {Class. Quant. Grav.}\ }\textbf {\bibinfo {volume}
  {40}},\ \bibinfo {pages} {195024} (\bibinfo {year} {2023})},\ \Eprint
  {http://arxiv.org/abs/2210.00295} {arXiv:2210.00295 [gr-qc]} \BibitemShut
  {NoStop}%
\bibitem [{\citenamefont {Franzin}\ \emph {et~al.}(2024)\citenamefont
  {Franzin}, \citenamefont {Liberati},\ and\ \citenamefont
  {Vellucci}}]{Franzin:2023slm}%
  \BibitemOpen
  \bibfield  {author} {\bibinfo {author} {\bibfnamefont {E.}~\bibnamefont
  {Franzin}}, \bibinfo {author} {\bibfnamefont {S.}~\bibnamefont {Liberati}}, \
  and\ \bibinfo {author} {\bibfnamefont {V.}~\bibnamefont {Vellucci}},\ }\href
  {\doibase 10.1088/1475-7516/2024/01/020} {\bibfield  {journal} {\bibinfo
  {journal} {JCAP}\ }\textbf {\bibinfo {volume} {01}},\ \bibinfo {pages} {020}
  (\bibinfo {year} {2024})},\ \Eprint {http://arxiv.org/abs/2310.11990}
  {arXiv:2310.11990 [gr-qc]} \BibitemShut {NoStop}%
\bibitem [{\citenamefont {Konoplya}\ \emph {et~al.}(2023)\citenamefont
  {Konoplya}, \citenamefont {Ovchinnikov},\ and\ \citenamefont
  {Ahmedov}}]{Konoplya:2023ahd}%
  \BibitemOpen
  \bibfield  {author} {\bibinfo {author} {\bibfnamefont {R.~A.}\ \bibnamefont
  {Konoplya}}, \bibinfo {author} {\bibfnamefont {D.}~\bibnamefont
  {Ovchinnikov}}, \ and\ \bibinfo {author} {\bibfnamefont {B.}~\bibnamefont
  {Ahmedov}},\ }\href {\doibase 10.1103/PhysRevD.108.104054} {\bibfield
  {journal} {\bibinfo  {journal} {Phys. Rev. D}\ }\textbf {\bibinfo {volume}
  {108}},\ \bibinfo {pages} {104054} (\bibinfo {year} {2023})},\ \Eprint
  {http://arxiv.org/abs/2307.10801} {arXiv:2307.10801 [gr-qc]} \BibitemShut
  {NoStop}%
\bibitem [{\citenamefont {Bolokhov}(2024{\natexlab{a}})}]{Bolokhov:2023ruj}%
  \BibitemOpen
  \bibfield  {author} {\bibinfo {author} {\bibfnamefont {S.~V.}\ \bibnamefont
  {Bolokhov}},\ }\href {\doibase 10.1103/PhysRevD.109.064017} {\bibfield
  {journal} {\bibinfo  {journal} {Phys. Rev. D}\ }\textbf {\bibinfo {volume}
  {109}},\ \bibinfo {pages} {064017} (\bibinfo {year}
  {2024}{\natexlab{a}})}\BibitemShut {NoStop}%
\bibitem [{\citenamefont {Konoplya}(2025)}]{Konoplya:2023bpf}%
  \BibitemOpen
  \bibfield  {author} {\bibinfo {author} {\bibfnamefont {R.~A.}\ \bibnamefont
  {Konoplya}},\ }\href {\doibase 10.1002/prop.202400002} {\bibfield  {journal}
  {\bibinfo  {journal} {Fortsch. Phys.}\ }\textbf {\bibinfo {volume} {73}},\
  \bibinfo {pages} {2400002} (\bibinfo {year} {2025})},\ \Eprint
  {http://arxiv.org/abs/2308.02850} {arXiv:2308.02850 [gr-qc]} \BibitemShut
  {NoStop}%
\bibitem [{\citenamefont {Skvortsova}(2025)}]{Skvortsova:2024eqi}%
  \BibitemOpen
  \bibfield  {author} {\bibinfo {author} {\bibfnamefont {M.}~\bibnamefont
  {Skvortsova}},\ }\href {\doibase 10.1209/0295-5075/adaee2} {\bibfield
  {journal} {\bibinfo  {journal} {EPL}\ }\textbf {\bibinfo {volume} {149}},\
  \bibinfo {pages} {59001} (\bibinfo {year} {2025})},\ \Eprint
  {http://arxiv.org/abs/2503.03650} {arXiv:2503.03650 [gr-qc]} \BibitemShut
  {NoStop}%
\bibitem [{\citenamefont {Pedrotti}\ and\ \citenamefont
  {Vagnozzi}(2024)}]{Pedrotti:2024znu}%
  \BibitemOpen
  \bibfield  {author} {\bibinfo {author} {\bibfnamefont {D.}~\bibnamefont
  {Pedrotti}}\ and\ \bibinfo {author} {\bibfnamefont {S.}~\bibnamefont
  {Vagnozzi}},\ }\href {\doibase 10.1103/PhysRevD.110.084075} {\bibfield
  {journal} {\bibinfo  {journal} {Phys. Rev. D}\ }\textbf {\bibinfo {volume}
  {110}},\ \bibinfo {pages} {084075} (\bibinfo {year} {2024})},\ \Eprint
  {http://arxiv.org/abs/2404.07589} {arXiv:2404.07589 [gr-qc]} \BibitemShut
  {NoStop}%
\bibitem [{\citenamefont {Stashko}(2024)}]{Stashko:2024wuq}%
  \BibitemOpen
  \bibfield  {author} {\bibinfo {author} {\bibfnamefont {O.}~\bibnamefont
  {Stashko}},\ }\href {\doibase 10.1103/PhysRevD.110.084016} {\bibfield
  {journal} {\bibinfo  {journal} {Phys. Rev. D}\ }\textbf {\bibinfo {volume}
  {110}},\ \bibinfo {pages} {084016} (\bibinfo {year} {2024})},\ \Eprint
  {http://arxiv.org/abs/2407.07892} {arXiv:2407.07892 [gr-qc]} \BibitemShut
  {NoStop}%
\bibitem [{\citenamefont {Skvortsova}(2024)}]{Skvortsova:2024wly}%
  \BibitemOpen
  \bibfield  {author} {\bibinfo {author} {\bibfnamefont {M.}~\bibnamefont
  {Skvortsova}},\ }\href {\doibase 10.1134/S020228932470018X} {\bibfield
  {journal} {\bibinfo  {journal} {Grav. Cosmol.}\ }\textbf {\bibinfo {volume}
  {30}},\ \bibinfo {pages} {279} (\bibinfo {year} {2024})},\ \Eprint
  {http://arxiv.org/abs/2405.15807} {arXiv:2405.15807 [gr-qc]} \BibitemShut
  {NoStop}%
\bibitem [{\citenamefont {Khoo}(2025)}]{Khoo:2025qjc}%
  \BibitemOpen
  \bibfield  {author} {\bibinfo {author} {\bibfnamefont {F.~S.}\ \bibnamefont
  {Khoo}},\ }\href {\doibase 10.1103/f35l-m8n5} {\bibfield  {journal} {\bibinfo
   {journal} {Phys. Rev. D}\ }\textbf {\bibinfo {volume} {111}},\ \bibinfo
  {pages} {124025} (\bibinfo {year} {2025})},\ \Eprint
  {http://arxiv.org/abs/2503.09390} {arXiv:2503.09390 [gr-qc]} \BibitemShut
  {NoStop}%
\bibitem [{\citenamefont {Konoplya}\ and\ \citenamefont
  {Zhidenko}(2025)}]{Konoplya:2025uta}%
  \BibitemOpen
  \bibfield  {author} {\bibinfo {author} {\bibfnamefont {R.~A.}\ \bibnamefont
  {Konoplya}}\ and\ \bibinfo {author} {\bibfnamefont {A.}~\bibnamefont
  {Zhidenko}},\ }\href {\doibase 10.53941/ijgtp.2025.100005} {\bibfield
  {journal} {\bibinfo  {journal} {Int. J. Grav. Theor. Phys.}\ }\textbf
  {\bibinfo {volume} {1}},\ \bibinfo {pages} {5} (\bibinfo {year} {2025})},\
  \Eprint {http://arxiv.org/abs/2507.22660} {arXiv:2507.22660 [gr-qc]}
  \BibitemShut {NoStop}%
\bibitem [{\citenamefont {Bolokhov}\ and\ \citenamefont
  {Skvortsova}(2025{\natexlab{a}})}]{Bolokhov:2025egl}%
  \BibitemOpen
  \bibfield  {author} {\bibinfo {author} {\bibfnamefont {S.~V.}\ \bibnamefont
  {Bolokhov}}\ and\ \bibinfo {author} {\bibfnamefont {M.}~\bibnamefont
  {Skvortsova}},\ }\href@noop {} {\  (\bibinfo {year} {2025}{\natexlab{a}})},\
  \Eprint {http://arxiv.org/abs/2508.19989} {arXiv:2508.19989 [gr-qc]}
  \BibitemShut {NoStop}%
\bibitem [{\citenamefont {Bolokhov}\ and\ \citenamefont
  {Skvortsova}(2025{\natexlab{b}})}]{Bolokhov:2025lnt}%
  \BibitemOpen
  \bibfield  {author} {\bibinfo {author} {\bibfnamefont {S.~V.}\ \bibnamefont
  {Bolokhov}}\ and\ \bibinfo {author} {\bibfnamefont {M.}~\bibnamefont
  {Skvortsova}},\ }\href {\doibase 10.53941/ijgtp.2025.100003} {\bibfield
  {journal} {\bibinfo  {journal} {Int. J. Grav. Theor. Phys.}\ }\textbf
  {\bibinfo {volume} {1}},\ \bibinfo {pages} {3} (\bibinfo {year}
  {2025}{\natexlab{b}})},\ \Eprint {http://arxiv.org/abs/2507.07196}
  {arXiv:2507.07196 [gr-qc]} \BibitemShut {NoStop}%
\bibitem [{\citenamefont {Arbelaez}(2025)}]{Arbelaez:2025gwj}%
  \BibitemOpen
  \bibfield  {author} {\bibinfo {author} {\bibfnamefont {J.~P.}\ \bibnamefont
  {Arbelaez}},\ }\href@noop {} {\  (\bibinfo {year} {2025})},\ \Eprint
  {http://arxiv.org/abs/2509.25141} {arXiv:2509.25141 [gr-qc]} \BibitemShut
  {NoStop}%
\bibitem [{\citenamefont {Navarro}\ \emph {et~al.}(1997)\citenamefont
  {Navarro}, \citenamefont {Frenk},\ and\ \citenamefont
  {White}}]{Navarro:1996gj}%
  \BibitemOpen
  \bibfield  {author} {\bibinfo {author} {\bibfnamefont {J.~F.}\ \bibnamefont
  {Navarro}}, \bibinfo {author} {\bibfnamefont {C.~S.}\ \bibnamefont {Frenk}},
  \ and\ \bibinfo {author} {\bibfnamefont {S.~D.~M.}\ \bibnamefont {White}},\
  }\href {\doibase 10.1086/304888} {\bibfield  {journal} {\bibinfo  {journal}
  {Astrophys. J.}\ }\textbf {\bibinfo {volume} {490}},\ \bibinfo {pages} {493}
  (\bibinfo {year} {1997})},\ \Eprint {http://arxiv.org/abs/astro-ph/9611107}
  {arXiv:astro-ph/9611107} \BibitemShut {NoStop}%
\bibitem [{\citenamefont {Bertone}\ \emph {et~al.}(2005)\citenamefont
  {Bertone}, \citenamefont {Hooper},\ and\ \citenamefont
  {Silk}}]{Bertone:2005xz}%
  \BibitemOpen
  \bibfield  {author} {\bibinfo {author} {\bibfnamefont {G.}~\bibnamefont
  {Bertone}}, \bibinfo {author} {\bibfnamefont {D.}~\bibnamefont {Hooper}}, \
  and\ \bibinfo {author} {\bibfnamefont {J.}~\bibnamefont {Silk}},\ }\href
  {\doibase 10.1016/j.physrep.2004.08.031} {\bibfield  {journal} {\bibinfo
  {journal} {Phys. Rept.}\ }\textbf {\bibinfo {volume} {405}},\ \bibinfo
  {pages} {279} (\bibinfo {year} {2005})},\ \Eprint
  {http://arxiv.org/abs/hep-ph/0404175} {arXiv:hep-ph/0404175} \BibitemShut
  {NoStop}%
\bibitem [{\citenamefont {Hernquist}(1990)}]{Hernquist:1990be}%
  \BibitemOpen
  \bibfield  {author} {\bibinfo {author} {\bibfnamefont {L.}~\bibnamefont
  {Hernquist}},\ }\href {\doibase 10.1086/168845} {\bibfield  {journal}
  {\bibinfo  {journal} {Astrophys. J.}\ }\textbf {\bibinfo {volume} {356}},\
  \bibinfo {pages} {359} (\bibinfo {year} {1990})}\BibitemShut {NoStop}%
\bibitem [{\citenamefont {Einasto}(1965{\natexlab{a}})}]{Einasto:1965czb}%
  \BibitemOpen
  \bibfield  {author} {\bibinfo {author} {\bibfnamefont {J.}~\bibnamefont
  {Einasto}},\ }\href@noop {} {\bibfield  {journal} {\bibinfo  {journal} {Trudy
  Astrofizicheskogo Instituta Alma-Ata}\ }\textbf {\bibinfo {volume} {5}},\
  \bibinfo {pages} {87} (\bibinfo {year} {1965}{\natexlab{a}})}\BibitemShut
  {NoStop}%
\bibitem [{\citenamefont {Dehnen}(1993)}]{Dehnen:1993uh}%
  \BibitemOpen
  \bibfield  {author} {\bibinfo {author} {\bibfnamefont {W.}~\bibnamefont
  {Dehnen}},\ }\href@noop {} {\bibfield  {journal} {\bibinfo  {journal} {Mon.
  Not. Roy. Astron. Soc.}\ }\textbf {\bibinfo {volume} {265}},\ \bibinfo
  {pages} {250} (\bibinfo {year} {1993})}\BibitemShut {NoStop}%
\bibitem [{\citenamefont {Dekel}\ \emph {et~al.}(2017)\citenamefont {Dekel},
  \citenamefont {Ishai}, \citenamefont {Dutton},\ and\ \citenamefont
  {Maccio}}]{Dekel:2017bwy}%
  \BibitemOpen
  \bibfield  {author} {\bibinfo {author} {\bibfnamefont {A.}~\bibnamefont
  {Dekel}}, \bibinfo {author} {\bibfnamefont {G.}~\bibnamefont {Ishai}},
  \bibinfo {author} {\bibfnamefont {A.~A.}\ \bibnamefont {Dutton}}, \ and\
  \bibinfo {author} {\bibfnamefont {A.~V.}\ \bibnamefont {Maccio}},\ }\href
  {\doibase 10.1093/mnras/stx486} {\bibfield  {journal} {\bibinfo  {journal}
  {Mon. Not. Roy. Astron. Soc.}\ }\textbf {\bibinfo {volume} {468}},\ \bibinfo
  {pages} {1005} (\bibinfo {year} {2017})},\ \Eprint
  {http://arxiv.org/abs/1610.00916} {arXiv:1610.00916 [astro-ph.GA]}
  \BibitemShut {NoStop}%
\bibitem [{\citenamefont {Zhao}(1996)}]{Zhao:1995cp}%
  \BibitemOpen
  \bibfield  {author} {\bibinfo {author} {\bibfnamefont {H.}~\bibnamefont
  {Zhao}},\ }\href {\doibase 10.1093/mnras/278.2.488} {\bibfield  {journal}
  {\bibinfo  {journal} {Mon. Not. Roy. Astron. Soc.}\ }\textbf {\bibinfo
  {volume} {278}},\ \bibinfo {pages} {488} (\bibinfo {year} {1996})},\ \Eprint
  {http://arxiv.org/abs/astro-ph/9509122} {arXiv:astro-ph/9509122} \BibitemShut
  {NoStop}%
\bibitem [{\citenamefont {Kar}\ and\ \citenamefont {Kar}(2025)}]{Kar:2025phe}%
  \BibitemOpen
  \bibfield  {author} {\bibinfo {author} {\bibfnamefont {A.}~\bibnamefont
  {Kar}}\ and\ \bibinfo {author} {\bibfnamefont {S.}~\bibnamefont {Kar}},\
  }\href {\doibase 10.1140/epjc/s10052-025-14483-5} {\bibfield  {journal}
  {\bibinfo  {journal} {Eur. Phys. J. C}\ }\textbf {\bibinfo {volume} {85}},\
  \bibinfo {pages} {773} (\bibinfo {year} {2025})},\ \Eprint
  {http://arxiv.org/abs/2504.12042} {arXiv:2504.12042 [gr-qc]} \BibitemShut
  {NoStop}%
\bibitem [{\citenamefont {Sajadi}\ \emph {et~al.}(2024)\citenamefont {Sajadi},
  \citenamefont {Khodadi}, \citenamefont {Luongo},\ and\ \citenamefont
  {Quevedo}}]{Sajadi:2023ybm}%
  \BibitemOpen
  \bibfield  {author} {\bibinfo {author} {\bibfnamefont {S.~N.}\ \bibnamefont
  {Sajadi}}, \bibinfo {author} {\bibfnamefont {M.}~\bibnamefont {Khodadi}},
  \bibinfo {author} {\bibfnamefont {O.}~\bibnamefont {Luongo}}, \ and\ \bibinfo
  {author} {\bibfnamefont {H.}~\bibnamefont {Quevedo}},\ }\href {\doibase
  10.1016/j.dark.2024.101525} {\bibfield  {journal} {\bibinfo  {journal} {Phys.
  Dark Univ.}\ }\textbf {\bibinfo {volume} {45}},\ \bibinfo {pages} {101525}
  (\bibinfo {year} {2024})},\ \Eprint {http://arxiv.org/abs/2312.16081}
  {arXiv:2312.16081 [gr-qc]} \BibitemShut {NoStop}%
\bibitem [{\citenamefont {Sajadi}\ \emph {et~al.}(2025)\citenamefont {Sajadi},
  \citenamefont {Ponglertsakul},\ and\ \citenamefont
  {Luongo}}]{Sajadi:2025prp}%
  \BibitemOpen
  \bibfield  {author} {\bibinfo {author} {\bibfnamefont {S.~N.}\ \bibnamefont
  {Sajadi}}, \bibinfo {author} {\bibfnamefont {S.}~\bibnamefont
  {Ponglertsakul}}, \ and\ \bibinfo {author} {\bibfnamefont {O.}~\bibnamefont
  {Luongo}},\ }\href {\doibase 10.1016/j.dark.2025.101938} {\bibfield
  {journal} {\bibinfo  {journal} {Phys. Dark Univ.}\ }\textbf {\bibinfo
  {volume} {48}},\ \bibinfo {pages} {101938} (\bibinfo {year} {2025})},\
  \Eprint {http://arxiv.org/abs/2502.02098} {arXiv:2502.02098 [gr-qc]}
  \BibitemShut {NoStop}%
\bibitem [{\citenamefont {Benson}(2010)}]{Benson:2010de}%
  \BibitemOpen
  \bibfield  {author} {\bibinfo {author} {\bibfnamefont {A.~J.}\ \bibnamefont
  {Benson}},\ }\href {\doibase 10.1016/j.physrep.2010.06.001} {\bibfield
  {journal} {\bibinfo  {journal} {Phys. Rept.}\ }\textbf {\bibinfo {volume}
  {495}},\ \bibinfo {pages} {33} (\bibinfo {year} {2010})},\ \Eprint
  {http://arxiv.org/abs/1006.5394} {arXiv:1006.5394 [astro-ph.CO]} \BibitemShut
  {NoStop}%
\bibitem [{\citenamefont {Konoplya}\ and\ \citenamefont
  {Zhidenko}(2022)}]{Konoplya:2022hbl}%
  \BibitemOpen
  \bibfield  {author} {\bibinfo {author} {\bibfnamefont {R.~A.}\ \bibnamefont
  {Konoplya}}\ and\ \bibinfo {author} {\bibfnamefont {A.}~\bibnamefont
  {Zhidenko}},\ }\href {\doibase 10.3847/1538-4357/ac76bc} {\bibfield
  {journal} {\bibinfo  {journal} {Astrophys. J.}\ }\textbf {\bibinfo {volume}
  {933}},\ \bibinfo {pages} {166} (\bibinfo {year} {2022})},\ \Eprint
  {http://arxiv.org/abs/2202.02205} {arXiv:2202.02205 [gr-qc]} \BibitemShut
  {NoStop}%
\bibitem [{\citenamefont {Zaslavskii}(2025)}]{Zaslavskii:2025oli}%
  \BibitemOpen
  \bibfield  {author} {\bibinfo {author} {\bibfnamefont {O.~B.}\ \bibnamefont
  {Zaslavskii}},\ }\href@noop {} {\  (\bibinfo {year} {2025})},\ \Eprint
  {http://arxiv.org/abs/2509.00899} {arXiv:2509.00899 [gr-qc]} \BibitemShut
  {NoStop}%
\bibitem [{\citenamefont {Kamenshchik}\ \emph {et~al.}(2001)\citenamefont
  {Kamenshchik}, \citenamefont {Moschella},\ and\ \citenamefont
  {Pasquier}}]{Kamenshchik:2001cp}%
  \BibitemOpen
  \bibfield  {author} {\bibinfo {author} {\bibfnamefont {A.~Y.}\ \bibnamefont
  {Kamenshchik}}, \bibinfo {author} {\bibfnamefont {U.}~\bibnamefont
  {Moschella}}, \ and\ \bibinfo {author} {\bibfnamefont {V.}~\bibnamefont
  {Pasquier}},\ }\href@noop {} {\bibfield  {journal} {\bibinfo  {journal}
  {Phys. Lett. B}\ }\textbf {\bibinfo {volume} {511}},\ \bibinfo {pages} {265}
  (\bibinfo {year} {2001})},\ \Eprint {http://arxiv.org/abs/gr-qc/0103004}
  {gr-qc/0103004} \BibitemShut {NoStop}%
\bibitem [{\citenamefont {Bento}\ \emph {et~al.}(2002)\citenamefont {Bento},
  \citenamefont {Bertolami},\ and\ \citenamefont {Sen}}]{Bento:2002ps}%
  \BibitemOpen
  \bibfield  {author} {\bibinfo {author} {\bibfnamefont {M.~C.}\ \bibnamefont
  {Bento}}, \bibinfo {author} {\bibfnamefont {O.}~\bibnamefont {Bertolami}}, \
  and\ \bibinfo {author} {\bibfnamefont {A.~A.}\ \bibnamefont {Sen}},\
  }\href@noop {} {\bibfield  {journal} {\bibinfo  {journal} {Phys. Rev. D}\
  }\textbf {\bibinfo {volume} {66}},\ \bibinfo {pages} {043507} (\bibinfo
  {year} {2002})},\ \Eprint {http://arxiv.org/abs/gr-qc/0202064}
  {gr-qc/0202064} \BibitemShut {NoStop}%
\bibitem [{\citenamefont {Scherrer}(2004)}]{Scherrer:2004au}%
  \BibitemOpen
  \bibfield  {author} {\bibinfo {author} {\bibfnamefont {R.~J.}\ \bibnamefont
  {Scherrer}},\ }\href@noop {} {\bibfield  {journal} {\bibinfo  {journal}
  {Phys. Rev. Lett.}\ }\textbf {\bibinfo {volume} {93}},\ \bibinfo {pages}
  {011301} (\bibinfo {year} {2004})},\ \Eprint
  {http://arxiv.org/abs/astro-ph/0402316} {astro-ph/0402316} \BibitemShut
  {NoStop}%
\bibitem [{\citenamefont {Dymnikova}\ and\ \citenamefont
  {Khlopov}(2015)}]{Dymnikova:2015yma}%
  \BibitemOpen
  \bibfield  {author} {\bibinfo {author} {\bibfnamefont {I.}~\bibnamefont
  {Dymnikova}}\ and\ \bibinfo {author} {\bibfnamefont {M.}~\bibnamefont
  {Khlopov}},\ }\href {\doibase 10.1142/S0218271815450029} {\bibfield
  {journal} {\bibinfo  {journal} {Int. J. Mod. Phys. D}\ }\textbf {\bibinfo
  {volume} {24}},\ \bibinfo {pages} {1545002} (\bibinfo {year} {2015})},\
  \Eprint {http://arxiv.org/abs/1510.01351} {arXiv:1510.01351 [gr-qc]}
  \BibitemShut {NoStop}%
\bibitem [{\citenamefont {Balakin}\ \emph {et~al.}(2003)\citenamefont
  {Balakin}, \citenamefont {Pavón}, \citenamefont {Schwarz},\ and\
  \citenamefont {Zimdahl}}]{Balakin:2003tk}%
  \BibitemOpen
  \bibfield  {author} {\bibinfo {author} {\bibfnamefont {A.~B.}\ \bibnamefont
  {Balakin}}, \bibinfo {author} {\bibfnamefont {D.}~\bibnamefont {Pavón}},
  \bibinfo {author} {\bibfnamefont {D.~J.}\ \bibnamefont {Schwarz}}, \ and\
  \bibinfo {author} {\bibfnamefont {W.}~\bibnamefont {Zimdahl}},\ }\href@noop
  {} {\bibfield  {journal} {\bibinfo  {journal} {New J. Phys.}\ }\textbf
  {\bibinfo {volume} {5}},\ \bibinfo {pages} {85} (\bibinfo {year} {2003})},\
  \Eprint {http://arxiv.org/abs/astro-ph/0302150} {astro-ph/0302150}
  \BibitemShut {NoStop}%
\bibitem [{\citenamefont {Sahni}\ and\ \citenamefont
  {Shtanov}(2003)}]{Sahni:2004ai}%
  \BibitemOpen
  \bibfield  {author} {\bibinfo {author} {\bibfnamefont {V.}~\bibnamefont
  {Sahni}}\ and\ \bibinfo {author} {\bibfnamefont {Y.}~\bibnamefont
  {Shtanov}},\ }\href@noop {} {\bibfield  {journal} {\bibinfo  {journal}
  {JCAP}\ }\textbf {\bibinfo {volume} {0311}},\ \bibinfo {pages} {014}
  (\bibinfo {year} {2003})},\ \Eprint {http://arxiv.org/abs/astro-ph/0202346}
  {astro-ph/0202346} \BibitemShut {NoStop}%
\bibitem [{\citenamefont {Einasto}(1965{\natexlab{b}})}]{Einasto:1965}%
  \BibitemOpen
  \bibfield  {author} {\bibinfo {author} {\bibfnamefont {J.}~\bibnamefont
  {Einasto}},\ }\href@noop {} {\bibfield  {journal} {\bibinfo  {journal} {Trudy
  Astrofizicheskogo Instituta Alma-Ata}\ }\textbf {\bibinfo {volume} {5}},\
  \bibinfo {pages} {87} (\bibinfo {year} {1965}{\natexlab{b}})},\ \bibinfo
  {note} {in Russian}\BibitemShut {NoStop}%
\bibitem [{\citenamefont {Einasto}\ and\ \citenamefont
  {Haud}(1989)}]{Einasto:1989}%
  \BibitemOpen
  \bibfield  {author} {\bibinfo {author} {\bibfnamefont {J.}~\bibnamefont
  {Einasto}}\ and\ \bibinfo {author} {\bibfnamefont {U.}~\bibnamefont {Haud}},\
  }\href@noop {} {\bibfield  {journal} {\bibinfo  {journal} {Astronomy and
  Astrophysics}\ }\textbf {\bibinfo {volume} {223}},\ \bibinfo {pages} {89}
  (\bibinfo {year} {1989})}\BibitemShut {NoStop}%
\bibitem [{\citenamefont {Retana-Montenegro}\ \emph {et~al.}(2012)\citenamefont
  {Retana-Montenegro}, \citenamefont {Van~Hese}, \citenamefont {Gentile},
  \citenamefont {Baes},\ and\ \citenamefont
  {Frutos-Alfaro}}]{Retana-Montenegro:2012dbd}%
  \BibitemOpen
  \bibfield  {author} {\bibinfo {author} {\bibfnamefont {E.}~\bibnamefont
  {Retana-Montenegro}}, \bibinfo {author} {\bibfnamefont {E.}~\bibnamefont
  {Van~Hese}}, \bibinfo {author} {\bibfnamefont {G.}~\bibnamefont {Gentile}},
  \bibinfo {author} {\bibfnamefont {M.}~\bibnamefont {Baes}}, \ and\ \bibinfo
  {author} {\bibfnamefont {F.}~\bibnamefont {Frutos-Alfaro}},\ }\href {\doibase
  10.1051/0004-6361/201118543} {\bibfield  {journal} {\bibinfo  {journal}
  {Astron. Astrophys.}\ }\textbf {\bibinfo {volume} {540}},\ \bibinfo {pages}
  {A70} (\bibinfo {year} {2012})},\ \Eprint {http://arxiv.org/abs/1202.5242}
  {arXiv:1202.5242 [astro-ph.CO]} \BibitemShut {NoStop}%
\bibitem [{\citenamefont {Baes}(2022)}]{Baes:2022pbc}%
  \BibitemOpen
  \bibfield  {author} {\bibinfo {author} {\bibfnamefont {M.}~\bibnamefont
  {Baes}},\ }\href {\doibase 10.1051/0004-6361/202244567} {\bibfield  {journal}
  {\bibinfo  {journal} {Astron. Astrophys.}\ }\textbf {\bibinfo {volume}
  {667}},\ \bibinfo {pages} {A47} (\bibinfo {year} {2022})},\ \Eprint
  {http://arxiv.org/abs/2209.03639} {arXiv:2209.03639 [astro-ph.GA]}
  \BibitemShut {NoStop}%
\bibitem [{\citenamefont {Merritt}\ \emph {et~al.}(2006)\citenamefont
  {Merritt}, \citenamefont {Navarro}, \citenamefont {Ludlow},\ and\
  \citenamefont {Jenkins}}]{Merritt:2006}%
  \BibitemOpen
  \bibfield  {author} {\bibinfo {author} {\bibfnamefont {D.}~\bibnamefont
  {Merritt}}, \bibinfo {author} {\bibfnamefont {J.~F.}\ \bibnamefont
  {Navarro}}, \bibinfo {author} {\bibfnamefont {A.}~\bibnamefont {Ludlow}}, \
  and\ \bibinfo {author} {\bibfnamefont {A.}~\bibnamefont {Jenkins}},\
  }\href@noop {} {\bibfield  {journal} {\bibinfo  {journal} {Astronomical
  Journal}\ }\textbf {\bibinfo {volume} {132}},\ \bibinfo {pages} {2685}
  (\bibinfo {year} {2006})}\BibitemShut {NoStop}%
\bibitem [{\citenamefont {Navarro}\ \emph {et~al.}(2004)\citenamefont {Navarro}
  \emph {et~al.}}]{MNRAS:Navarro2004}%
  \BibitemOpen
  \bibfield  {author} {\bibinfo {author} {\bibfnamefont {J.~F.}\ \bibnamefont
  {Navarro}} \emph {et~al.},\ }\href@noop {} {\bibfield  {journal} {\bibinfo
  {journal} {Monthly Notices of the Royal Astronomical Society}\ }\textbf
  {\bibinfo {volume} {349}},\ \bibinfo {pages} {1039} (\bibinfo {year}
  {2004})}\BibitemShut {NoStop}%
\bibitem [{\citenamefont {Springel}\ \emph {et~al.}(2008)\citenamefont
  {Springel} \emph {et~al.}}]{Springel:2008}%
  \BibitemOpen
  \bibfield  {author} {\bibinfo {author} {\bibfnamefont {V.}~\bibnamefont
  {Springel}} \emph {et~al.},\ }\href@noop {} {\bibfield  {journal} {\bibinfo
  {journal} {Monthly Notices of the Royal Astronomical Society}\ }\textbf
  {\bibinfo {volume} {391}},\ \bibinfo {pages} {1685} (\bibinfo {year}
  {2008})}\BibitemShut {NoStop}%
\bibitem [{\citenamefont {Acharyya}\ \emph {et~al.}(2024)\citenamefont
  {Acharyya}, \citenamefont {Banerjee},\ and\ \citenamefont
  {Kar}}]{Acharyya:2023rnq}%
  \BibitemOpen
  \bibfield  {author} {\bibinfo {author} {\bibfnamefont {R.}~\bibnamefont
  {Acharyya}}, \bibinfo {author} {\bibfnamefont {P.}~\bibnamefont {Banerjee}},
  \ and\ \bibinfo {author} {\bibfnamefont {S.}~\bibnamefont {Kar}},\ }\href
  {\doibase 10.1088/1475-7516/2024/04/070} {\bibfield  {journal} {\bibinfo
  {journal} {JCAP}\ }\textbf {\bibinfo {volume} {04}},\ \bibinfo {pages} {070}
  (\bibinfo {year} {2024})},\ \Eprint {http://arxiv.org/abs/2311.18622}
  {arXiv:2311.18622 [gr-qc]} \BibitemShut {NoStop}%
\bibitem [{\citenamefont {Graham}\ \emph {et~al.}(2006)\citenamefont {Graham},
  \citenamefont {Merritt}, \citenamefont {Moore}, \citenamefont {Diemand},\
  and\ \citenamefont {Terzić}}]{Graham:2006}%
  \BibitemOpen
  \bibfield  {author} {\bibinfo {author} {\bibfnamefont {A.~W.}\ \bibnamefont
  {Graham}}, \bibinfo {author} {\bibfnamefont {D.}~\bibnamefont {Merritt}},
  \bibinfo {author} {\bibfnamefont {B.}~\bibnamefont {Moore}}, \bibinfo
  {author} {\bibfnamefont {J.}~\bibnamefont {Diemand}}, \ and\ \bibinfo
  {author} {\bibfnamefont {B.}~\bibnamefont {Terzić}},\ }\href@noop {}
  {\bibfield  {journal} {\bibinfo  {journal} {Astrophysical Journal Letters}\
  }\textbf {\bibinfo {volume} {653}},\ \bibinfo {pages} {L121} (\bibinfo {year}
  {2006})}\BibitemShut {NoStop}%
\bibitem [{\citenamefont {Ludlow}\ \emph {et~al.}(2013)\citenamefont {Ludlow},
  \citenamefont {Navarro}, \citenamefont {Angulo} \emph
  {et~al.}}]{Ludlow:2013}%
  \BibitemOpen
  \bibfield  {author} {\bibinfo {author} {\bibfnamefont {A.~D.}\ \bibnamefont
  {Ludlow}}, \bibinfo {author} {\bibfnamefont {J.~F.}\ \bibnamefont {Navarro}},
  \bibinfo {author} {\bibfnamefont {R.~E.}\ \bibnamefont {Angulo}},  \emph
  {et~al.},\ }\href@noop {} {\bibfield  {journal} {\bibinfo  {journal} {Monthly
  Notices of the Royal Astronomical Society}\ }\textbf {\bibinfo {volume}
  {432}},\ \bibinfo {pages} {1103} (\bibinfo {year} {2013})}\BibitemShut
  {NoStop}%
\bibitem [{\citenamefont {Boos}(2025)}]{Boos:2021kqe}%
  \BibitemOpen
  \bibfield  {author} {\bibinfo {author} {\bibfnamefont {J.}~\bibnamefont
  {Boos}},\ }\href {\doibase 10.3390/universe11040112} {\bibfield  {journal}
  {\bibinfo  {journal} {Universe}\ }\textbf {\bibinfo {volume} {11}},\ \bibinfo
  {pages} {112} (\bibinfo {year} {2025})},\ \Eprint
  {http://arxiv.org/abs/2104.00555} {arXiv:2104.00555 [gr-qc]} \BibitemShut
  {NoStop}%
\bibitem [{\citenamefont {Taylor}\ and\ \citenamefont
  {Silk}(2003)}]{Taylor:2002zd}%
  \BibitemOpen
  \bibfield  {author} {\bibinfo {author} {\bibfnamefont {J.~E.}\ \bibnamefont
  {Taylor}}\ and\ \bibinfo {author} {\bibfnamefont {J.}~\bibnamefont {Silk}},\
  }\href {\doibase 10.1046/j.1365-8711.2003.06201.x} {\bibfield  {journal}
  {\bibinfo  {journal} {Mon. Not. Roy. Astron. Soc.}\ }\textbf {\bibinfo
  {volume} {339}},\ \bibinfo {pages} {505} (\bibinfo {year} {2003})},\ \Eprint
  {http://arxiv.org/abs/astro-ph/0207299} {arXiv:astro-ph/0207299} \BibitemShut
  {NoStop}%
\bibitem [{\citenamefont {Moore}\ \emph {et~al.}(1998)\citenamefont {Moore},
  \citenamefont {Governato}, \citenamefont {Quinn}, \citenamefont {Stadel},\
  and\ \citenamefont {Lake}}]{Moore:1997sg}%
  \BibitemOpen
  \bibfield  {author} {\bibinfo {author} {\bibfnamefont {B.}~\bibnamefont
  {Moore}}, \bibinfo {author} {\bibfnamefont {F.}~\bibnamefont {Governato}},
  \bibinfo {author} {\bibfnamefont {T.~R.}\ \bibnamefont {Quinn}}, \bibinfo
  {author} {\bibfnamefont {J.}~\bibnamefont {Stadel}}, \ and\ \bibinfo {author}
  {\bibfnamefont {G.}~\bibnamefont {Lake}},\ }\href {\doibase 10.1086/311333}
  {\bibfield  {journal} {\bibinfo  {journal} {Astrophys. J. Lett.}\ }\textbf
  {\bibinfo {volume} {499}},\ \bibinfo {pages} {L5} (\bibinfo {year} {1998})},\
  \Eprint {http://arxiv.org/abs/astro-ph/9709051} {arXiv:astro-ph/9709051}
  \BibitemShut {NoStop}%
\bibitem [{\citenamefont {Dymnikova}(2004)}]{Dymnikova:2004zc}%
  \BibitemOpen
  \bibfield  {author} {\bibinfo {author} {\bibfnamefont {I.}~\bibnamefont
  {Dymnikova}},\ }\href {\doibase 10.1088/0264-9381/21/18/009} {\bibfield
  {journal} {\bibinfo  {journal} {Class. Quant. Grav.}\ }\textbf {\bibinfo
  {volume} {21}},\ \bibinfo {pages} {4417} (\bibinfo {year} {2004})},\ \Eprint
  {http://arxiv.org/abs/gr-qc/0407072} {arXiv:gr-qc/0407072} \BibitemShut
  {NoStop}%
\bibitem [{\citenamefont {Chakraborty}\ \emph {et~al.}(2025)\citenamefont
  {Chakraborty}, \citenamefont {Compère},\ and\ \citenamefont
  {Machet}}]{Chakraborty:2024gcr}%
  \BibitemOpen
  \bibfield  {author} {\bibinfo {author} {\bibfnamefont {S.}~\bibnamefont
  {Chakraborty}}, \bibinfo {author} {\bibfnamefont {G.}~\bibnamefont
  {Compère}}, \ and\ \bibinfo {author} {\bibfnamefont {L.}~\bibnamefont
  {Machet}},\ }\href {\doibase 10.1103/4p2c-rwdh} {\bibfield  {journal}
  {\bibinfo  {journal} {Phys. Rev. D}\ }\textbf {\bibinfo {volume} {112}},\
  \bibinfo {pages} {024015} (\bibinfo {year} {2025})},\ \Eprint
  {http://arxiv.org/abs/2412.14831} {arXiv:2412.14831 [gr-qc]} \BibitemShut
  {NoStop}%
\bibitem [{\citenamefont {Konoplya}\ and\ \citenamefont
  {Zhidenko}(2011)}]{Konoplya:2011qq}%
  \BibitemOpen
  \bibfield  {author} {\bibinfo {author} {\bibfnamefont {R.~A.}\ \bibnamefont
  {Konoplya}}\ and\ \bibinfo {author} {\bibfnamefont {A.}~\bibnamefont
  {Zhidenko}},\ }\href {\doibase 10.1103/RevModPhys.83.793} {\bibfield
  {journal} {\bibinfo  {journal} {Rev. Mod. Phys.}\ }\textbf {\bibinfo {volume}
  {83}},\ \bibinfo {pages} {793} (\bibinfo {year} {2011})},\ \Eprint
  {http://arxiv.org/abs/1102.4014} {arXiv:1102.4014 [gr-qc]} \BibitemShut
  {NoStop}%
\bibitem [{\citenamefont {Cunha}\ and\ \citenamefont
  {Herdeiro}(2018)}]{Cunha:2018acu}%
  \BibitemOpen
  \bibfield  {author} {\bibinfo {author} {\bibfnamefont {P.~V.~P.}\
  \bibnamefont {Cunha}}\ and\ \bibinfo {author} {\bibfnamefont {C.~A.~R.}\
  \bibnamefont {Herdeiro}},\ }\href {\doibase 10.1007/s10714-018-2361-9}
  {\bibfield  {journal} {\bibinfo  {journal} {Gen. Rel. Grav.}\ }\textbf
  {\bibinfo {volume} {50}},\ \bibinfo {pages} {42} (\bibinfo {year} {2018})},\
  \Eprint {http://arxiv.org/abs/1801.00860} {arXiv:1801.00860 [gr-qc]}
  \BibitemShut {NoStop}%
\bibitem [{\citenamefont {Perlick}\ and\ \citenamefont
  {Tsupko}(2022)}]{Perlick:2021aok}%
  \BibitemOpen
  \bibfield  {author} {\bibinfo {author} {\bibfnamefont {V.}~\bibnamefont
  {Perlick}}\ and\ \bibinfo {author} {\bibfnamefont {O.~Y.}\ \bibnamefont
  {Tsupko}},\ }\href {\doibase 10.1016/j.physrep.2021.10.004} {\bibfield
  {journal} {\bibinfo  {journal} {Phys. Rept.}\ }\textbf {\bibinfo {volume}
  {947}},\ \bibinfo {pages} {1} (\bibinfo {year} {2022})},\ \Eprint
  {http://arxiv.org/abs/2105.07101} {arXiv:2105.07101 [gr-qc]} \BibitemShut
  {NoStop}%
\bibitem [{\citenamefont {Bambi}(2017)}]{Bambi:2015kza}%
  \BibitemOpen
  \bibfield  {author} {\bibinfo {author} {\bibfnamefont {C.}~\bibnamefont
  {Bambi}},\ }\href {\doibase 10.1103/RevModPhys.89.025001} {\bibfield
  {journal} {\bibinfo  {journal} {Rev. Mod. Phys.}\ }\textbf {\bibinfo {volume}
  {89}},\ \bibinfo {pages} {025001} (\bibinfo {year} {2017})},\ \Eprint
  {http://arxiv.org/abs/1509.03884} {arXiv:1509.03884 [gr-qc]} \BibitemShut
  {NoStop}%
\bibitem [{\citenamefont {Hioki}\ and\ \citenamefont
  {Maeda}(2009)}]{Hioki:2009na}%
  \BibitemOpen
  \bibfield  {author} {\bibinfo {author} {\bibfnamefont {K.}~\bibnamefont
  {Hioki}}\ and\ \bibinfo {author} {\bibfnamefont {K.-i.}\ \bibnamefont
  {Maeda}},\ }\href {\doibase 10.1103/PhysRevD.80.024042} {\bibfield  {journal}
  {\bibinfo  {journal} {Phys. Rev. D}\ }\textbf {\bibinfo {volume} {80}},\
  \bibinfo {pages} {024042} (\bibinfo {year} {2009})},\ \Eprint
  {http://arxiv.org/abs/0904.3575} {arXiv:0904.3575 [astro-ph.HE]} \BibitemShut
  {NoStop}%
\bibitem [{\citenamefont {Konoplya}(2019)}]{Konoplya:2019sns}%
  \BibitemOpen
  \bibfield  {author} {\bibinfo {author} {\bibfnamefont {R.~A.}\ \bibnamefont
  {Konoplya}},\ }\href {\doibase 10.1016/j.physletb.2019.05.043} {\bibfield
  {journal} {\bibinfo  {journal} {Phys. Lett. B}\ }\textbf {\bibinfo {volume}
  {795}},\ \bibinfo {pages} {1} (\bibinfo {year} {2019})},\ \Eprint
  {http://arxiv.org/abs/1905.00064} {arXiv:1905.00064 [gr-qc]} \BibitemShut
  {NoStop}%
\bibitem [{\citenamefont {Younsi}\ \emph {et~al.}(2016)\citenamefont {Younsi},
  \citenamefont {Zhidenko}, \citenamefont {Rezzolla}, \citenamefont
  {Konoplya},\ and\ \citenamefont {Mizuno}}]{Younsi:2016azx}%
  \BibitemOpen
  \bibfield  {author} {\bibinfo {author} {\bibfnamefont {Z.}~\bibnamefont
  {Younsi}}, \bibinfo {author} {\bibfnamefont {A.}~\bibnamefont {Zhidenko}},
  \bibinfo {author} {\bibfnamefont {L.}~\bibnamefont {Rezzolla}}, \bibinfo
  {author} {\bibfnamefont {R.}~\bibnamefont {Konoplya}}, \ and\ \bibinfo
  {author} {\bibfnamefont {Y.}~\bibnamefont {Mizuno}},\ }\href {\doibase
  10.1103/PhysRevD.94.084025} {\bibfield  {journal} {\bibinfo  {journal} {Phys.
  Rev. D}\ }\textbf {\bibinfo {volume} {94}},\ \bibinfo {pages} {084025}
  (\bibinfo {year} {2016})},\ \Eprint {http://arxiv.org/abs/1607.05767}
  {arXiv:1607.05767 [gr-qc]} \BibitemShut {NoStop}%
\bibitem [{\citenamefont {Bambi}\ \emph {et~al.}(2019)\citenamefont {Bambi},
  \citenamefont {Freese}, \citenamefont {Vagnozzi},\ and\ \citenamefont
  {Visinelli}}]{Bambi:2019tjh}%
  \BibitemOpen
  \bibfield  {author} {\bibinfo {author} {\bibfnamefont {C.}~\bibnamefont
  {Bambi}}, \bibinfo {author} {\bibfnamefont {K.}~\bibnamefont {Freese}},
  \bibinfo {author} {\bibfnamefont {S.}~\bibnamefont {Vagnozzi}}, \ and\
  \bibinfo {author} {\bibfnamefont {L.}~\bibnamefont {Visinelli}},\ }\href
  {\doibase 10.1103/PhysRevD.100.044057} {\bibfield  {journal} {\bibinfo
  {journal} {Phys. Rev. D}\ }\textbf {\bibinfo {volume} {100}},\ \bibinfo
  {pages} {044057} (\bibinfo {year} {2019})},\ \Eprint
  {http://arxiv.org/abs/1904.12983} {arXiv:1904.12983 [gr-qc]} \BibitemShut
  {NoStop}%
\bibitem [{\citenamefont {Konoplya}\ \emph
  {et~al.}(2025{\natexlab{a}})\citenamefont {Konoplya}, \citenamefont
  {Ovchinnikov},\ and\ \citenamefont {Schee}}]{Konoplya:2025bte}%
  \BibitemOpen
  \bibfield  {author} {\bibinfo {author} {\bibfnamefont {R.~A.}\ \bibnamefont
  {Konoplya}}, \bibinfo {author} {\bibfnamefont {D.}~\bibnamefont
  {Ovchinnikov}}, \ and\ \bibinfo {author} {\bibfnamefont {J.}~\bibnamefont
  {Schee}},\ }\href@noop {} {\  (\bibinfo {year} {2025}{\natexlab{a}})},\
  \Eprint {http://arxiv.org/abs/2510.05947} {arXiv:2510.05947 [gr-qc]}
  \BibitemShut {NoStop}%
\bibitem [{\citenamefont {Perlick}\ \emph {et~al.}(2015)\citenamefont
  {Perlick}, \citenamefont {Tsupko},\ and\ \citenamefont
  {Bisnovatyi-Kogan}}]{Perlick:2015vta}%
  \BibitemOpen
  \bibfield  {author} {\bibinfo {author} {\bibfnamefont {V.}~\bibnamefont
  {Perlick}}, \bibinfo {author} {\bibfnamefont {O.~Y.}\ \bibnamefont {Tsupko}},
  \ and\ \bibinfo {author} {\bibfnamefont {G.~S.}\ \bibnamefont
  {Bisnovatyi-Kogan}},\ }\href {\doibase 10.1103/PhysRevD.92.104031} {\bibfield
   {journal} {\bibinfo  {journal} {Phys. Rev. D}\ }\textbf {\bibinfo {volume}
  {92}},\ \bibinfo {pages} {104031} (\bibinfo {year} {2015})},\ \Eprint
  {http://arxiv.org/abs/1507.04217} {arXiv:1507.04217 [gr-qc]} \BibitemShut
  {NoStop}%
\bibitem [{\citenamefont {Konoplya}\ and\ \citenamefont
  {Stashko}(2025)}]{Konoplya:2024lch}%
  \BibitemOpen
  \bibfield  {author} {\bibinfo {author} {\bibfnamefont {R.~A.}\ \bibnamefont
  {Konoplya}}\ and\ \bibinfo {author} {\bibfnamefont {O.~S.}\ \bibnamefont
  {Stashko}},\ }\href {\doibase 10.1103/PhysRevD.111.104055} {\bibfield
  {journal} {\bibinfo  {journal} {Phys. Rev. D}\ }\textbf {\bibinfo {volume}
  {111}},\ \bibinfo {pages} {104055} (\bibinfo {year} {2025})},\ \Eprint
  {http://arxiv.org/abs/2408.02578} {arXiv:2408.02578 [gr-qc]} \BibitemShut
  {NoStop}%
\bibitem [{\citenamefont {Khodadi}\ \emph {et~al.}(2020)\citenamefont
  {Khodadi}, \citenamefont {Allahyari}, \citenamefont {Vagnozzi},\ and\
  \citenamefont {Mota}}]{Khodadi:2020jij}%
  \BibitemOpen
  \bibfield  {author} {\bibinfo {author} {\bibfnamefont {M.}~\bibnamefont
  {Khodadi}}, \bibinfo {author} {\bibfnamefont {A.}~\bibnamefont {Allahyari}},
  \bibinfo {author} {\bibfnamefont {S.}~\bibnamefont {Vagnozzi}}, \ and\
  \bibinfo {author} {\bibfnamefont {D.~F.}\ \bibnamefont {Mota}},\ }\href
  {\doibase 10.1088/1475-7516/2020/09/026} {\bibfield  {journal} {\bibinfo
  {journal} {JCAP}\ }\textbf {\bibinfo {volume} {09}},\ \bibinfo {pages} {026}
  (\bibinfo {year} {2020})},\ \Eprint {http://arxiv.org/abs/2005.05992}
  {arXiv:2005.05992 [gr-qc]} \BibitemShut {NoStop}%
\bibitem [{\citenamefont {Kumar}\ and\ \citenamefont
  {Ghosh}(2020)}]{Kumar:2018ple}%
  \BibitemOpen
  \bibfield  {author} {\bibinfo {author} {\bibfnamefont {R.}~\bibnamefont
  {Kumar}}\ and\ \bibinfo {author} {\bibfnamefont {S.~G.}\ \bibnamefont
  {Ghosh}},\ }\href {\doibase 10.3847/1538-4357/ab77b0} {\bibfield  {journal}
  {\bibinfo  {journal} {Astrophys. J.}\ }\textbf {\bibinfo {volume} {892}},\
  \bibinfo {pages} {78} (\bibinfo {year} {2020})},\ \Eprint
  {http://arxiv.org/abs/1811.01260} {arXiv:1811.01260 [gr-qc]} \BibitemShut
  {NoStop}%
\bibitem [{\citenamefont {Konoplya}\ and\ \citenamefont
  {Zhidenko}(2021)}]{Konoplya:2021slg}%
  \BibitemOpen
  \bibfield  {author} {\bibinfo {author} {\bibfnamefont {R.~A.}\ \bibnamefont
  {Konoplya}}\ and\ \bibinfo {author} {\bibfnamefont {A.}~\bibnamefont
  {Zhidenko}},\ }\href {\doibase 10.1103/PhysRevD.103.104033} {\bibfield
  {journal} {\bibinfo  {journal} {Phys. Rev. D}\ }\textbf {\bibinfo {volume}
  {103}},\ \bibinfo {pages} {104033} (\bibinfo {year} {2021})},\ \Eprint
  {http://arxiv.org/abs/2103.03855} {arXiv:2103.03855 [gr-qc]} \BibitemShut
  {NoStop}%
\bibitem [{\citenamefont {Afrin}\ \emph {et~al.}(2021)\citenamefont {Afrin},
  \citenamefont {Kumar},\ and\ \citenamefont {Ghosh}}]{Afrin:2021imp}%
  \BibitemOpen
  \bibfield  {author} {\bibinfo {author} {\bibfnamefont {M.}~\bibnamefont
  {Afrin}}, \bibinfo {author} {\bibfnamefont {R.}~\bibnamefont {Kumar}}, \ and\
  \bibinfo {author} {\bibfnamefont {S.~G.}\ \bibnamefont {Ghosh}},\ }\href
  {\doibase 10.1093/mnras/stab1260} {\bibfield  {journal} {\bibinfo  {journal}
  {Mon. Not. Roy. Astron. Soc.}\ }\textbf {\bibinfo {volume} {504}},\ \bibinfo
  {pages} {5927} (\bibinfo {year} {2021})},\ \Eprint
  {http://arxiv.org/abs/2103.11417} {arXiv:2103.11417 [gr-qc]} \BibitemShut
  {NoStop}%
\bibitem [{\citenamefont {Zakharov}(2014)}]{Zakharov:2014lqa}%
  \BibitemOpen
  \bibfield  {author} {\bibinfo {author} {\bibfnamefont {A.~F.}\ \bibnamefont
  {Zakharov}},\ }\href {\doibase 10.1103/PhysRevD.90.062007} {\bibfield
  {journal} {\bibinfo  {journal} {Phys. Rev. D}\ }\textbf {\bibinfo {volume}
  {90}},\ \bibinfo {pages} {062007} (\bibinfo {year} {2014})},\ \Eprint
  {http://arxiv.org/abs/1407.7457} {arXiv:1407.7457 [gr-qc]} \BibitemShut
  {NoStop}%
\bibitem [{\citenamefont {Zakharov}(2025{\natexlab{a}})}]{Zakharov:2025rhb}%
  \BibitemOpen
  \bibfield  {author} {\bibinfo {author} {\bibfnamefont {A.~F.}\ \bibnamefont
  {Zakharov}},\ }\href {\doibase 10.54546/NaturalSciRev.100402} {\bibfield
  {journal} {\bibinfo  {journal} {Natural Sci. Rev.}\ }\textbf {\bibinfo
  {volume} {2}},\ \bibinfo {pages} {100402} (\bibinfo {year}
  {2025}{\natexlab{a}})},\ \Eprint {http://arxiv.org/abs/2506.16927}
  {arXiv:2506.16927 [physics.hist-ph]} \BibitemShut {NoStop}%
\bibitem [{\citenamefont {Zakharov}(2025{\natexlab{b}})}]{Zakharov:2025cnq}%
  \BibitemOpen
  \bibfield  {author} {\bibinfo {author} {\bibfnamefont {A.~F.}\ \bibnamefont
  {Zakharov}},\ }\href {\doibase 10.1134/S106377882570019X} {\bibfield
  {journal} {\bibinfo  {journal} {Phys. Atom. Nucl.}\ }\textbf {\bibinfo
  {volume} {88}},\ \bibinfo {pages} {154} (\bibinfo {year}
  {2025}{\natexlab{b}})}\BibitemShut {NoStop}%
\bibitem [{\citenamefont
  {Lütfüoğlu}(2025{\natexlab{a}})}]{Lutfuoglu:2025ldc}%
  \BibitemOpen
  \bibfield  {author} {\bibinfo {author} {\bibfnamefont {B.~C.}\ \bibnamefont
  {Lütfüoğlu}},\ }\href {\doibase 10.53941/ijgtp.2025.100004} {\bibfield
  {journal} {\bibinfo  {journal} {Int. J. Grav. Theor. Phys.}\ }\textbf
  {\bibinfo {volume} {1}},\ \bibinfo {pages} {4} (\bibinfo {year}
  {2025}{\natexlab{a}})},\ \Eprint {http://arxiv.org/abs/2507.09246}
  {arXiv:2507.09246 [gr-qc]} \BibitemShut {NoStop}%
\bibitem [{\citenamefont {Konoplya}\ \emph {et~al.}(2020)\citenamefont
  {Konoplya}, \citenamefont {Pappas},\ and\ \citenamefont
  {Zhidenko}}]{Konoplya:2019fpy}%
  \BibitemOpen
  \bibfield  {author} {\bibinfo {author} {\bibfnamefont {R.~A.}\ \bibnamefont
  {Konoplya}}, \bibinfo {author} {\bibfnamefont {T.}~\bibnamefont {Pappas}}, \
  and\ \bibinfo {author} {\bibfnamefont {A.}~\bibnamefont {Zhidenko}},\ }\href
  {\doibase 10.1103/PhysRevD.101.044054} {\bibfield  {journal} {\bibinfo
  {journal} {Phys. Rev. D}\ }\textbf {\bibinfo {volume} {101}},\ \bibinfo
  {pages} {044054} (\bibinfo {year} {2020})},\ \Eprint
  {http://arxiv.org/abs/1907.10112} {arXiv:1907.10112 [gr-qc]} \BibitemShut
  {NoStop}%
\bibitem [{\citenamefont {Konoplya}\ and\ \citenamefont
  {Zhidenko}(2019)}]{Konoplya:2019goy}%
  \BibitemOpen
  \bibfield  {author} {\bibinfo {author} {\bibfnamefont {R.~A.}\ \bibnamefont
  {Konoplya}}\ and\ \bibinfo {author} {\bibfnamefont {A.}~\bibnamefont
  {Zhidenko}},\ }\href {\doibase 10.1103/PhysRevD.100.044015} {\bibfield
  {journal} {\bibinfo  {journal} {Phys. Rev. D}\ }\textbf {\bibinfo {volume}
  {100}},\ \bibinfo {pages} {044015} (\bibinfo {year} {2019})},\ \Eprint
  {http://arxiv.org/abs/1907.05551} {arXiv:1907.05551 [gr-qc]} \BibitemShut
  {NoStop}%
\bibitem [{\citenamefont {Tsukamoto}(2018)}]{Tsukamoto:2017fxq}%
  \BibitemOpen
  \bibfield  {author} {\bibinfo {author} {\bibfnamefont {N.}~\bibnamefont
  {Tsukamoto}},\ }\href {\doibase 10.1103/PhysRevD.97.064021} {\bibfield
  {journal} {\bibinfo  {journal} {Phys. Rev. D}\ }\textbf {\bibinfo {volume}
  {97}},\ \bibinfo {pages} {064021} (\bibinfo {year} {2018})},\ \Eprint
  {http://arxiv.org/abs/1708.07427} {arXiv:1708.07427 [gr-qc]} \BibitemShut
  {NoStop}%
\bibitem [{\citenamefont {Cardoso}\ \emph {et~al.}(2009)\citenamefont
  {Cardoso}, \citenamefont {Miranda}, \citenamefont {Berti}, \citenamefont
  {Witek},\ and\ \citenamefont {Zanchin}}]{Cardoso:2008bp}%
  \BibitemOpen
  \bibfield  {author} {\bibinfo {author} {\bibfnamefont {V.}~\bibnamefont
  {Cardoso}}, \bibinfo {author} {\bibfnamefont {A.~S.}\ \bibnamefont
  {Miranda}}, \bibinfo {author} {\bibfnamefont {E.}~\bibnamefont {Berti}},
  \bibinfo {author} {\bibfnamefont {H.}~\bibnamefont {Witek}}, \ and\ \bibinfo
  {author} {\bibfnamefont {V.~T.}\ \bibnamefont {Zanchin}},\ }\href {\doibase
  10.1103/PhysRevD.79.064016} {\bibfield  {journal} {\bibinfo  {journal} {Phys.
  Rev. D}\ }\textbf {\bibinfo {volume} {79}},\ \bibinfo {pages} {064016}
  (\bibinfo {year} {2009})},\ \Eprint {http://arxiv.org/abs/0812.1806}
  {arXiv:0812.1806 [hep-th]} \BibitemShut {NoStop}%
\bibitem [{\citenamefont {Jusufi}(2020)}]{Jusufi:2020dhz}%
  \BibitemOpen
  \bibfield  {author} {\bibinfo {author} {\bibfnamefont {K.}~\bibnamefont
  {Jusufi}},\ }\href {\doibase 10.1103/PhysRevD.101.124063} {\bibfield
  {journal} {\bibinfo  {journal} {Phys. Rev. D}\ }\textbf {\bibinfo {volume}
  {101}},\ \bibinfo {pages} {124063} (\bibinfo {year} {2020})},\ \Eprint
  {http://arxiv.org/abs/2004.04664} {arXiv:2004.04664 [gr-qc]} \BibitemShut
  {NoStop}%
\bibitem [{\citenamefont {Konoplya}\ and\ \citenamefont
  {Stuchlík}(2017)}]{Konoplya:2017wot}%
  \BibitemOpen
  \bibfield  {author} {\bibinfo {author} {\bibfnamefont {R.~A.}\ \bibnamefont
  {Konoplya}}\ and\ \bibinfo {author} {\bibfnamefont {Z.}~\bibnamefont
  {Stuchlík}},\ }\href {\doibase 10.1016/j.physletb.2017.06.015} {\bibfield
  {journal} {\bibinfo  {journal} {Phys. Lett. B}\ }\textbf {\bibinfo {volume}
  {771}},\ \bibinfo {pages} {597} (\bibinfo {year} {2017})},\ \Eprint
  {http://arxiv.org/abs/1705.05928} {arXiv:1705.05928 [gr-qc]} \BibitemShut
  {NoStop}%
\bibitem [{\citenamefont {Khanna}\ and\ \citenamefont
  {Price}(2017)}]{Khanna:2016yow}%
  \BibitemOpen
  \bibfield  {author} {\bibinfo {author} {\bibfnamefont {G.}~\bibnamefont
  {Khanna}}\ and\ \bibinfo {author} {\bibfnamefont {R.~H.}\ \bibnamefont
  {Price}},\ }\href {\doibase 10.1103/PhysRevD.95.081501} {\bibfield  {journal}
  {\bibinfo  {journal} {Phys. Rev. D}\ }\textbf {\bibinfo {volume} {95}},\
  \bibinfo {pages} {081501} (\bibinfo {year} {2017})},\ \Eprint
  {http://arxiv.org/abs/1609.00083} {arXiv:1609.00083 [gr-qc]} \BibitemShut
  {NoStop}%
\bibitem [{\citenamefont {Konoplya}(2023)}]{Konoplya:2022gjp}%
  \BibitemOpen
  \bibfield  {author} {\bibinfo {author} {\bibfnamefont {R.~A.}\ \bibnamefont
  {Konoplya}},\ }\href {\doibase 10.1016/j.physletb.2023.137674} {\bibfield
  {journal} {\bibinfo  {journal} {Phys. Lett. B}\ }\textbf {\bibinfo {volume}
  {838}},\ \bibinfo {pages} {137674} (\bibinfo {year} {2023})},\ \Eprint
  {http://arxiv.org/abs/2210.08373} {arXiv:2210.08373 [gr-qc]} \BibitemShut
  {NoStop}%
\bibitem [{\citenamefont {Bolokhov}(2024{\natexlab{b}})}]{Bolokhov:2023dxq}%
  \BibitemOpen
  \bibfield  {author} {\bibinfo {author} {\bibfnamefont {S.~V.}\ \bibnamefont
  {Bolokhov}},\ }\href {\doibase 10.1016/j.physletb.2024.138879} {\bibfield
  {journal} {\bibinfo  {journal} {Phys. Lett. B}\ }\textbf {\bibinfo {volume}
  {856}},\ \bibinfo {pages} {138879} (\bibinfo {year} {2024}{\natexlab{b}})},\
  \Eprint {http://arxiv.org/abs/2310.12326} {arXiv:2310.12326 [gr-qc]}
  \BibitemShut {NoStop}%
\bibitem [{\citenamefont {Konoplya}\ and\ \citenamefont
  {Zinhailo}(2020)}]{Konoplya:2020bxa}%
  \BibitemOpen
  \bibfield  {author} {\bibinfo {author} {\bibfnamefont {R.~A.}\ \bibnamefont
  {Konoplya}}\ and\ \bibinfo {author} {\bibfnamefont {A.~F.}\ \bibnamefont
  {Zinhailo}},\ }\href {\doibase 10.1140/epjc/s10052-020-08639-8} {\bibfield
  {journal} {\bibinfo  {journal} {Eur. Phys. J. C}\ }\textbf {\bibinfo {volume}
  {80}},\ \bibinfo {pages} {1049} (\bibinfo {year} {2020})},\ \Eprint
  {http://arxiv.org/abs/2003.01188} {arXiv:2003.01188 [gr-qc]} \BibitemShut
  {NoStop}%
\bibitem [{\citenamefont {Konoplya}\ \emph
  {et~al.}(2025{\natexlab{b}})\citenamefont {Konoplya}, \citenamefont {Spina},\
  and\ \citenamefont {Zhidenko}}]{Konoplya:2025afm}%
  \BibitemOpen
  \bibfield  {author} {\bibinfo {author} {\bibfnamefont {R.~A.}\ \bibnamefont
  {Konoplya}}, \bibinfo {author} {\bibfnamefont {A.}~\bibnamefont {Spina}}, \
  and\ \bibinfo {author} {\bibfnamefont {A.}~\bibnamefont {Zhidenko}},\ }\href
  {\doibase 10.1103/xhtc-9cf4} {\bibfield  {journal} {\bibinfo  {journal}
  {Phys. Rev. D}\ }\textbf {\bibinfo {volume} {112}},\ \bibinfo {pages}
  {024060} (\bibinfo {year} {2025}{\natexlab{b}})},\ \Eprint
  {http://arxiv.org/abs/2505.01128} {arXiv:2505.01128 [gr-qc]} \BibitemShut
  {NoStop}%
\bibitem [{\citenamefont {Huang}\ and\ \citenamefont
  {Rao}(2025)}]{Huang:2025uhv}%
  \BibitemOpen
  \bibfield  {author} {\bibinfo {author} {\bibfnamefont {H.}~\bibnamefont
  {Huang}}\ and\ \bibinfo {author} {\bibfnamefont {X.-P.}\ \bibnamefont
  {Rao}},\ }\href {\doibase 10.1103/PhysRevD.111.104040} {\bibfield  {journal}
  {\bibinfo  {journal} {Phys. Rev. D}\ }\textbf {\bibinfo {volume} {111}},\
  \bibinfo {pages} {104040} (\bibinfo {year} {2025})},\ \Eprint
  {http://arxiv.org/abs/2503.13133} {arXiv:2503.13133 [gr-qc]} \BibitemShut
  {NoStop}%
\bibitem [{\citenamefont {Datta}(2024)}]{Datta:2023zmd}%
  \BibitemOpen
  \bibfield  {author} {\bibinfo {author} {\bibfnamefont {S.}~\bibnamefont
  {Datta}},\ }\href {\doibase 10.1103/PhysRevD.109.104042} {\bibfield
  {journal} {\bibinfo  {journal} {Phys. Rev. D}\ }\textbf {\bibinfo {volume}
  {109}},\ \bibinfo {pages} {104042} (\bibinfo {year} {2024})},\ \Eprint
  {http://arxiv.org/abs/2312.01277} {arXiv:2312.01277 [gr-qc]} \BibitemShut
  {NoStop}%
\bibitem [{\citenamefont {Springel}\ \emph {et~al.}(2005)\citenamefont
  {Springel} \emph {et~al.}}]{Springel:2005nw}%
  \BibitemOpen
  \bibfield  {author} {\bibinfo {author} {\bibfnamefont {V.}~\bibnamefont
  {Springel}} \emph {et~al.},\ }\href {\doibase 10.1038/nature03597} {\bibfield
   {journal} {\bibinfo  {journal} {Nature}\ }\textbf {\bibinfo {volume}
  {435}},\ \bibinfo {pages} {629} (\bibinfo {year} {2005})},\ \Eprint
  {http://arxiv.org/abs/astro-ph/0504097} {arXiv:astro-ph/0504097} \BibitemShut
  {NoStop}%
\bibitem [{\citenamefont {Spergel}\ and\ \citenamefont
  {Steinhardt}(2000)}]{Spergel:1999mh}%
  \BibitemOpen
  \bibfield  {author} {\bibinfo {author} {\bibfnamefont {D.~N.}\ \bibnamefont
  {Spergel}}\ and\ \bibinfo {author} {\bibfnamefont {P.~J.}\ \bibnamefont
  {Steinhardt}},\ }\href {\doibase 10.1103/PhysRevLett.84.3760} {\bibfield
  {journal} {\bibinfo  {journal} {Phys. Rev. Lett.}\ }\textbf {\bibinfo
  {volume} {84}},\ \bibinfo {pages} {3760} (\bibinfo {year} {2000})},\ \Eprint
  {http://arxiv.org/abs/astro-ph/9909386} {arXiv:astro-ph/9909386} \BibitemShut
  {NoStop}%
\bibitem [{\citenamefont {Tulin}\ and\ \citenamefont
  {Yu}(2018)}]{Tulin:2017ara}%
  \BibitemOpen
  \bibfield  {author} {\bibinfo {author} {\bibfnamefont {S.}~\bibnamefont
  {Tulin}}\ and\ \bibinfo {author} {\bibfnamefont {H.-B.}\ \bibnamefont {Yu}},\
  }\href {\doibase 10.1016/j.physrep.2017.11.004} {\bibfield  {journal}
  {\bibinfo  {journal} {Phys. Rept.}\ }\textbf {\bibinfo {volume} {730}},\
  \bibinfo {pages} {1} (\bibinfo {year} {2018})},\ \Eprint
  {http://arxiv.org/abs/1705.02358} {arXiv:1705.02358 [hep-ph]} \BibitemShut
  {NoStop}%
\bibitem [{\citenamefont {Gondolo}\ and\ \citenamefont
  {Silk}(1999)}]{Gondolo:1999ef}%
  \BibitemOpen
  \bibfield  {author} {\bibinfo {author} {\bibfnamefont {P.}~\bibnamefont
  {Gondolo}}\ and\ \bibinfo {author} {\bibfnamefont {J.}~\bibnamefont {Silk}},\
  }\href {\doibase 10.1103/PhysRevLett.83.1719} {\bibfield  {journal} {\bibinfo
   {journal} {Phys. Rev. Lett.}\ }\textbf {\bibinfo {volume} {83}},\ \bibinfo
  {pages} {1719} (\bibinfo {year} {1999})},\ \Eprint
  {http://arxiv.org/abs/astro-ph/9906391} {arXiv:astro-ph/9906391} \BibitemShut
  {NoStop}%
\bibitem [{\citenamefont {Ullio}\ \emph {et~al.}(2001)\citenamefont {Ullio},
  \citenamefont {Zhao},\ and\ \citenamefont {Kamionkowski}}]{Ullio:2001fb}%
  \BibitemOpen
  \bibfield  {author} {\bibinfo {author} {\bibfnamefont {P.}~\bibnamefont
  {Ullio}}, \bibinfo {author} {\bibfnamefont {H.}~\bibnamefont {Zhao}}, \ and\
  \bibinfo {author} {\bibfnamefont {M.}~\bibnamefont {Kamionkowski}},\ }\href
  {\doibase 10.1103/PhysRevD.64.043504} {\bibfield  {journal} {\bibinfo
  {journal} {Phys. Rev. D}\ }\textbf {\bibinfo {volume} {64}},\ \bibinfo
  {pages} {043504} (\bibinfo {year} {2001})},\ \Eprint
  {http://arxiv.org/abs/astro-ph/0101481} {arXiv:astro-ph/0101481} \BibitemShut
  {NoStop}%
\bibitem [{\citenamefont {Dialektopoulos}\ \emph {et~al.}(2025)\citenamefont
  {Dialektopoulos}, \citenamefont {Papanikolaou},\ and\ \citenamefont
  {Zarikas}}]{Dialektopoulos:2025mfz}%
  \BibitemOpen
  \bibfield  {author} {\bibinfo {author} {\bibfnamefont {K.}~\bibnamefont
  {Dialektopoulos}}, \bibinfo {author} {\bibfnamefont {T.}~\bibnamefont
  {Papanikolaou}}, \ and\ \bibinfo {author} {\bibfnamefont {V.}~\bibnamefont
  {Zarikas}},\ }\href {\doibase 10.1016/j.physletb.2025.139948} {\bibfield
  {journal} {\bibinfo  {journal} {Phys. Lett. B}\ }\textbf {\bibinfo {volume}
  {870}},\ \bibinfo {pages} {139948} (\bibinfo {year} {2025})},\ \Eprint
  {http://arxiv.org/abs/2502.18352} {arXiv:2502.18352 [gr-qc]} \BibitemShut
  {NoStop}%
\bibitem [{\citenamefont {Vagnozzi}\ \emph {et~al.}(2023)\citenamefont
  {Vagnozzi} \emph {et~al.}}]{Vagnozzi:2022moj}%
  \BibitemOpen
  \bibfield  {author} {\bibinfo {author} {\bibfnamefont {S.}~\bibnamefont
  {Vagnozzi}} \emph {et~al.},\ }\href {\doibase 10.1088/1361-6382/acd97b}
  {\bibfield  {journal} {\bibinfo  {journal} {Class. Quant. Grav.}\ }\textbf
  {\bibinfo {volume} {40}},\ \bibinfo {pages} {165007} (\bibinfo {year}
  {2023})},\ \Eprint {http://arxiv.org/abs/2205.07787} {arXiv:2205.07787
  [gr-qc]} \BibitemShut {NoStop}%
\bibitem [{\citenamefont {Boos}\ and\ \citenamefont {Hu}(2025)}]{Boos:2025nzc}%
  \BibitemOpen
  \bibfield  {author} {\bibinfo {author} {\bibfnamefont {J.}~\bibnamefont
  {Boos}}\ and\ \bibinfo {author} {\bibfnamefont {H.}~\bibnamefont {Hu}},\
  }\href@noop {} {\  (\bibinfo {year} {2025})},\ \Eprint
  {http://arxiv.org/abs/2510.10282} {arXiv:2510.10282 [gr-qc]} \BibitemShut
  {NoStop}%
\bibitem [{\citenamefont {Konoplya}(2021)}]{Konoplya:2021ube}%
  \BibitemOpen
  \bibfield  {author} {\bibinfo {author} {\bibfnamefont {R.~A.}\ \bibnamefont
  {Konoplya}},\ }\href {\doibase 10.1016/j.physletb.2021.136734} {\bibfield
  {journal} {\bibinfo  {journal} {Phys. Lett. B}\ }\textbf {\bibinfo {volume}
  {823}},\ \bibinfo {pages} {136734} (\bibinfo {year} {2021})},\ \Eprint
  {http://arxiv.org/abs/2109.01640} {arXiv:2109.01640 [gr-qc]} \BibitemShut
  {NoStop}%
\bibitem [{\citenamefont {Dubinsky}(2025)}]{Dubinsky:2025fwv}%
  \BibitemOpen
  \bibfield  {author} {\bibinfo {author} {\bibfnamefont {A.}~\bibnamefont
  {Dubinsky}},\ }\href@noop {} {\bibfield  {journal} {\bibinfo  {journal} {Int.
  J. Grav. Theor. Phys.}\ }\textbf {\bibinfo {volume} {1}},\ \bibinfo {pages}
  {2} (\bibinfo {year} {2025})},\ \Eprint {http://arxiv.org/abs/2507.00256}
  {arXiv:2507.00256 [gr-qc]} \BibitemShut {NoStop}%
\bibitem [{\citenamefont {Feng}\ and\ \citenamefont
  {Zhang}(2025)}]{Feng:2025iao}%
  \BibitemOpen
  \bibfield  {author} {\bibinfo {author} {\bibfnamefont {X.-H.}\ \bibnamefont
  {Feng}}\ and\ \bibinfo {author} {\bibfnamefont {G.-Y.}\ \bibnamefont
  {Zhang}},\ }\href@noop {} {\  (\bibinfo {year} {2025})},\ \Eprint
  {http://arxiv.org/abs/2509.04001} {arXiv:2509.04001 [gr-qc]} \BibitemShut
  {NoStop}%
\bibitem [{\citenamefont {Pezzella}\ \emph {et~al.}(2025)\citenamefont
  {Pezzella}, \citenamefont {Destounis}, \citenamefont {Maselli},\ and\
  \citenamefont {Cardoso}}]{Pezzella:2024tkf}%
  \BibitemOpen
  \bibfield  {author} {\bibinfo {author} {\bibfnamefont {L.}~\bibnamefont
  {Pezzella}}, \bibinfo {author} {\bibfnamefont {K.}~\bibnamefont {Destounis}},
  \bibinfo {author} {\bibfnamefont {A.}~\bibnamefont {Maselli}}, \ and\
  \bibinfo {author} {\bibfnamefont {V.}~\bibnamefont {Cardoso}},\ }\href
  {\doibase 10.1103/PhysRevD.111.064026} {\bibfield  {journal} {\bibinfo
  {journal} {Phys. Rev. D}\ }\textbf {\bibinfo {volume} {111}},\ \bibinfo
  {pages} {064026} (\bibinfo {year} {2025})},\ \Eprint
  {http://arxiv.org/abs/2412.18651} {arXiv:2412.18651 [gr-qc]} \BibitemShut
  {NoStop}%
\bibitem [{\citenamefont {Liu}\ \emph {et~al.}(2025)\citenamefont {Liu},
  \citenamefont {Mu}, \citenamefont {Tao},\ and\ \citenamefont
  {Weng}}]{Liu:2024bfj}%
  \BibitemOpen
  \bibfield  {author} {\bibinfo {author} {\bibfnamefont {Y.}~\bibnamefont
  {Liu}}, \bibinfo {author} {\bibfnamefont {B.}~\bibnamefont {Mu}}, \bibinfo
  {author} {\bibfnamefont {J.}~\bibnamefont {Tao}}, \ and\ \bibinfo {author}
  {\bibfnamefont {Y.}~\bibnamefont {Weng}},\ }\href {\doibase
  10.1016/j.nuclphysb.2024.116787} {\bibfield  {journal} {\bibinfo  {journal}
  {Nucl. Phys. B}\ }\textbf {\bibinfo {volume} {1010}},\ \bibinfo {pages}
  {116787} (\bibinfo {year} {2025})},\ \Eprint
  {http://arxiv.org/abs/2409.20333} {arXiv:2409.20333 [gr-qc]} \BibitemShut
  {NoStop}%
\bibitem [{\citenamefont {Liu}\ \emph {et~al.}(2024)\citenamefont {Liu},
  \citenamefont {Yang},\ and\ \citenamefont {Long}}]{Liu:2024xcd}%
  \BibitemOpen
  \bibfield  {author} {\bibinfo {author} {\bibfnamefont {D.}~\bibnamefont
  {Liu}}, \bibinfo {author} {\bibfnamefont {Y.}~\bibnamefont {Yang}}, \ and\
  \bibinfo {author} {\bibfnamefont {Z.-W.}\ \bibnamefont {Long}},\ }\href
  {\doibase 10.1140/epjc/s10052-024-13096-8} {\bibfield  {journal} {\bibinfo
  {journal} {Eur. Phys. J. C}\ }\textbf {\bibinfo {volume} {84}},\ \bibinfo
  {pages} {731} (\bibinfo {year} {2024})},\ \Eprint
  {http://arxiv.org/abs/2401.09182} {arXiv:2401.09182 [gr-qc]} \BibitemShut
  {NoStop}%
\bibitem [{\citenamefont {Zhao}\ \emph {et~al.}(2023)\citenamefont {Zhao},
  \citenamefont {Sun}, \citenamefont {Lin},\ and\ \citenamefont
  {Cao}}]{Zhao:2023tyo}%
  \BibitemOpen
  \bibfield  {author} {\bibinfo {author} {\bibfnamefont {Y.}~\bibnamefont
  {Zhao}}, \bibinfo {author} {\bibfnamefont {B.}~\bibnamefont {Sun}}, \bibinfo
  {author} {\bibfnamefont {K.}~\bibnamefont {Lin}}, \ and\ \bibinfo {author}
  {\bibfnamefont {Z.}~\bibnamefont {Cao}},\ }\href {\doibase
  10.1103/PhysRevD.108.024070} {\bibfield  {journal} {\bibinfo  {journal}
  {Phys. Rev. D}\ }\textbf {\bibinfo {volume} {108}},\ \bibinfo {pages}
  {024070} (\bibinfo {year} {2023})},\ \Eprint
  {http://arxiv.org/abs/2303.09215} {arXiv:2303.09215 [gr-qc]} \BibitemShut
  {NoStop}%
\bibitem [{\citenamefont {Daghigh}\ and\ \citenamefont
  {Kunstatter}(2022)}]{Daghigh:2022pcr}%
  \BibitemOpen
  \bibfield  {author} {\bibinfo {author} {\bibfnamefont {R.~G.}\ \bibnamefont
  {Daghigh}}\ and\ \bibinfo {author} {\bibfnamefont {G.}~\bibnamefont
  {Kunstatter}},\ }\href {\doibase 10.3847/1538-4357/ac940b} {\bibfield
  {journal} {\bibinfo  {journal} {Astrophys. J.}\ }\textbf {\bibinfo {volume}
  {940}},\ \bibinfo {pages} {33} (\bibinfo {year} {2022})},\ \bibinfo {note}
  {[Erratum: Astrophys.J. 963, 167 (2024)]},\ \Eprint
  {http://arxiv.org/abs/2206.04195} {arXiv:2206.04195 [astro-ph.GA]}
  \BibitemShut {NoStop}%
\bibitem [{\citenamefont {Zhang}\ \emph {et~al.}(2021)\citenamefont {Zhang},
  \citenamefont {Zhu},\ and\ \citenamefont {Wang}}]{Zhang:2021bdr}%
  \BibitemOpen
  \bibfield  {author} {\bibinfo {author} {\bibfnamefont {C.}~\bibnamefont
  {Zhang}}, \bibinfo {author} {\bibfnamefont {T.}~\bibnamefont {Zhu}}, \ and\
  \bibinfo {author} {\bibfnamefont {A.}~\bibnamefont {Wang}},\ }\href {\doibase
  10.1103/PhysRevD.104.124082} {\bibfield  {journal} {\bibinfo  {journal}
  {Phys. Rev. D}\ }\textbf {\bibinfo {volume} {104}},\ \bibinfo {pages}
  {124082} (\bibinfo {year} {2021})},\ \Eprint
  {http://arxiv.org/abs/2111.04966} {arXiv:2111.04966 [gr-qc]} \BibitemShut
  {NoStop}%
\bibitem [{\citenamefont {Hamil}\ \emph {et~al.}(2025)\citenamefont {Hamil},
  \citenamefont {Al-Badawi},\ and\ \citenamefont
  {Lütfüoğlu}}]{Hamil:2025pte}%
  \BibitemOpen
  \bibfield  {author} {\bibinfo {author} {\bibfnamefont {B.}~\bibnamefont
  {Hamil}}, \bibinfo {author} {\bibfnamefont {A.}~\bibnamefont {Al-Badawi}}, \
  and\ \bibinfo {author} {\bibfnamefont {B.~C.}\ \bibnamefont {Lütfüoğlu}},\
  }\href@noop {} {\  (\bibinfo {year} {2025})},\ \Eprint
  {http://arxiv.org/abs/2505.18611} {arXiv:2505.18611 [gr-qc]} \BibitemShut
  {NoStop}%
\bibitem [{\citenamefont {Mollicone}\ and\ \citenamefont
  {Destounis}(2025)}]{Mollicone:2024lxy}%
  \BibitemOpen
  \bibfield  {author} {\bibinfo {author} {\bibfnamefont {A.}~\bibnamefont
  {Mollicone}}\ and\ \bibinfo {author} {\bibfnamefont {K.}~\bibnamefont
  {Destounis}},\ }\href {\doibase 10.1103/PhysRevD.111.024017} {\bibfield
  {journal} {\bibinfo  {journal} {Phys. Rev. D}\ }\textbf {\bibinfo {volume}
  {111}},\ \bibinfo {pages} {024017} (\bibinfo {year} {2025})},\ \Eprint
  {http://arxiv.org/abs/2410.11952} {arXiv:2410.11952 [gr-qc]} \BibitemShut
  {NoStop}%
\bibitem [{\citenamefont {Tovar}\ \emph {et~al.}(2025)\citenamefont {Tovar},
  \citenamefont {Pedraza}, \citenamefont {López},\ and\ \citenamefont
  {Arceo}}]{Tovar:2025apz}%
  \BibitemOpen
  \bibfield  {author} {\bibinfo {author} {\bibfnamefont {L.~O.~T.}\
  \bibnamefont {Tovar}}, \bibinfo {author} {\bibfnamefont {O.}~\bibnamefont
  {Pedraza}}, \bibinfo {author} {\bibfnamefont {L.~A.}\ \bibnamefont {López}},
  \ and\ \bibinfo {author} {\bibfnamefont {R.}~\bibnamefont {Arceo}},\
  }\href@noop {} {\  (\bibinfo {year} {2025})},\ \Eprint
  {http://arxiv.org/abs/2507.03627} {arXiv:2507.03627 [gr-qc]} \BibitemShut
  {NoStop}%
\bibitem [{\citenamefont
  {Lütfüoğlu}(2025{\natexlab{b}})}]{Lutfuoglu:2025kqp}%
  \BibitemOpen
  \bibfield  {author} {\bibinfo {author} {\bibfnamefont {B.~C.}\ \bibnamefont
  {Lütfüoğlu}},\ }\href@noop {} {\  (\bibinfo {year}
  {2025}{\natexlab{b}})},\ \Eprint {http://arxiv.org/abs/2510.25969}
  {arXiv:2510.25969 [gr-qc]} \BibitemShut {NoStop}%
\bibitem [{\citenamefont {Pathrikar}(2025)}]{Pathrikar:2025sin}%
  \BibitemOpen
  \bibfield  {author} {\bibinfo {author} {\bibfnamefont {A.}~\bibnamefont
  {Pathrikar}},\ }\href@noop {} {\  (\bibinfo {year} {2025})},\ \Eprint
  {http://arxiv.org/abs/2511.02355} {arXiv:2511.02355 [gr-qc]} \BibitemShut
  {NoStop}%
\bibitem [{\citenamefont {Konoplya}\ \emph
  {et~al.}(2025{\natexlab{c}})\citenamefont {Konoplya}, \citenamefont
  {Stuchlík},\ and\ \citenamefont {Zhidenko}}]{Konoplya:2025nqv}%
  \BibitemOpen
  \bibfield  {author} {\bibinfo {author} {\bibfnamefont {R.~A.}\ \bibnamefont
  {Konoplya}}, \bibinfo {author} {\bibfnamefont {Z.}~\bibnamefont {Stuchlík}},
  \ and\ \bibinfo {author} {\bibfnamefont {A.}~\bibnamefont {Zhidenko}},\
  }\href {\doibase 10.1103/l33g-6hlr} {\bibfield  {journal} {\bibinfo
  {journal} {Phys. Rev. D}\ }\textbf {\bibinfo {volume} {112}},\ \bibinfo
  {pages} {083014} (\bibinfo {year} {2025}{\natexlab{c}})},\ \Eprint
  {http://arxiv.org/abs/2509.03301} {arXiv:2509.03301 [gr-qc]} \BibitemShut
  {NoStop}%
\bibitem [{\citenamefont {Hou}\ \emph {et~al.}(2018)\citenamefont {Hou},
  \citenamefont {Xu},\ and\ \citenamefont {Wang}}]{Hou:2018avu}%
  \BibitemOpen
  \bibfield  {author} {\bibinfo {author} {\bibfnamefont {X.}~\bibnamefont
  {Hou}}, \bibinfo {author} {\bibfnamefont {Z.}~\bibnamefont {Xu}}, \ and\
  \bibinfo {author} {\bibfnamefont {J.}~\bibnamefont {Wang}},\ }\href {\doibase
  10.1088/1475-7516/2018/12/040} {\bibfield  {journal} {\bibinfo  {journal}
  {JCAP}\ }\textbf {\bibinfo {volume} {12}},\ \bibinfo {pages} {040} (\bibinfo
  {year} {2018})},\ \Eprint {http://arxiv.org/abs/1810.06381} {arXiv:1810.06381
  [gr-qc]} \BibitemShut {NoStop}%
\bibitem [{\citenamefont {Kouniatalis}\ \emph {et~al.}(2025)\citenamefont
  {Kouniatalis}, \citenamefont {Suvorov},\ and\ \citenamefont
  {Destounis}}]{Kouniatalis:2025itj}%
  \BibitemOpen
  \bibfield  {author} {\bibinfo {author} {\bibfnamefont {G.}~\bibnamefont
  {Kouniatalis}}, \bibinfo {author} {\bibfnamefont {A.~G.}\ \bibnamefont
  {Suvorov}}, \ and\ \bibinfo {author} {\bibfnamefont {K.}~\bibnamefont
  {Destounis}},\ }\href@noop {} {\  (\bibinfo {year} {2025})},\ \Eprint
  {http://arxiv.org/abs/2508.19333} {arXiv:2508.19333 [gr-qc]} \BibitemShut
  {NoStop}%
\bibitem [{\citenamefont {Fernandes}\ and\ \citenamefont
  {Cardoso}(2025)}]{Fernandes:2025osu}%
  \BibitemOpen
  \bibfield  {author} {\bibinfo {author} {\bibfnamefont {P.~G.~S.}\
  \bibnamefont {Fernandes}}\ and\ \bibinfo {author} {\bibfnamefont
  {V.}~\bibnamefont {Cardoso}},\ }\href@noop {} {\  (\bibinfo {year} {2025})},\
  \Eprint {http://arxiv.org/abs/2507.04389} {arXiv:2507.04389 [gr-qc]}
  \BibitemShut {NoStop}%
\bibitem [{\citenamefont {Chen}\ \emph {et~al.}(2024)\citenamefont {Chen},
  \citenamefont {Javed}, \citenamefont {Mustafa}, \citenamefont {Maurya},\ and\
  \citenamefont {Ray}}]{Chen:2024lpd}%
  \BibitemOpen
  \bibfield  {author} {\bibinfo {author} {\bibfnamefont {R.-Y.}\ \bibnamefont
  {Chen}}, \bibinfo {author} {\bibfnamefont {F.}~\bibnamefont {Javed}},
  \bibinfo {author} {\bibfnamefont {D.~G.}\ \bibnamefont {Mustafa}}, \bibinfo
  {author} {\bibfnamefont {S.~K.}\ \bibnamefont {Maurya}}, \ and\ \bibinfo
  {author} {\bibfnamefont {S.}~\bibnamefont {Ray}},\ }\href {\doibase
  10.1016/j.jheap.2024.09.010} {\bibfield  {journal} {\bibinfo  {journal}
  {JHEAp}\ }\textbf {\bibinfo {volume} {44}},\ \bibinfo {pages} {172} (\bibinfo
  {year} {2024})}\BibitemShut {NoStop}%
\bibitem [{\citenamefont {Tan}\ \emph {et~al.}(2025)\citenamefont {Tan},
  \citenamefont {Deng}, \citenamefont {Long},\ and\ \citenamefont
  {Jing}}]{Tan:2024hzw}%
  \BibitemOpen
  \bibfield  {author} {\bibinfo {author} {\bibfnamefont {Q.}~\bibnamefont
  {Tan}}, \bibinfo {author} {\bibfnamefont {W.}~\bibnamefont {Deng}}, \bibinfo
  {author} {\bibfnamefont {S.}~\bibnamefont {Long}}, \ and\ \bibinfo {author}
  {\bibfnamefont {J.}~\bibnamefont {Jing}},\ }\href {\doibase
  10.1088/1475-7516/2025/05/044} {\bibfield  {journal} {\bibinfo  {journal}
  {JCAP}\ }\textbf {\bibinfo {volume} {05}},\ \bibinfo {pages} {044} (\bibinfo
  {year} {2025})},\ \Eprint {http://arxiv.org/abs/2409.17760} {arXiv:2409.17760
  [gr-qc]} \BibitemShut {NoStop}%
\bibitem [{\citenamefont {Macedo}\ \emph {et~al.}(2024)\citenamefont {Macedo},
  \citenamefont {Rosa},\ and\ \citenamefont {Rubiera-Garcia}}]{Macedo:2024qky}%
  \BibitemOpen
  \bibfield  {author} {\bibinfo {author} {\bibfnamefont {C.~F.~B.}\
  \bibnamefont {Macedo}}, \bibinfo {author} {\bibfnamefont {J.~L.}\
  \bibnamefont {Rosa}}, \ and\ \bibinfo {author} {\bibfnamefont
  {D.}~\bibnamefont {Rubiera-Garcia}},\ }\href {\doibase
  10.1088/1475-7516/2024/07/046} {\bibfield  {journal} {\bibinfo  {journal}
  {JCAP}\ }\textbf {\bibinfo {volume} {07}},\ \bibinfo {pages} {046} (\bibinfo
  {year} {2024})},\ \Eprint {http://arxiv.org/abs/2402.13047} {arXiv:2402.13047
  [gr-qc]} \BibitemShut {NoStop}%
\bibitem [{\citenamefont {Xavier}\ \emph {et~al.}(2023)\citenamefont {Xavier},
  \citenamefont {Lima},\ and\ \citenamefont {Crispino}}]{Xavier:2023exm}%
  \BibitemOpen
  \bibfield  {author} {\bibinfo {author} {\bibfnamefont {S.~V. M. C.~B.}\
  \bibnamefont {Xavier}}, \bibinfo {author} {\bibfnamefont {H.~C.~D.}\
  \bibnamefont {Lima}, \bibfnamefont {Junior.}}, \ and\ \bibinfo {author}
  {\bibfnamefont {L.~C.~B.}\ \bibnamefont {Crispino}},\ }\href {\doibase
  10.1103/PhysRevD.107.064040} {\bibfield  {journal} {\bibinfo  {journal}
  {Phys. Rev. D}\ }\textbf {\bibinfo {volume} {107}},\ \bibinfo {pages}
  {064040} (\bibinfo {year} {2023})},\ \Eprint
  {http://arxiv.org/abs/2303.17666} {arXiv:2303.17666 [gr-qc]} \BibitemShut
  {NoStop}%
\bibitem [{\citenamefont {Figueiredo}\ \emph {et~al.}(2023)\citenamefont
  {Figueiredo}, \citenamefont {Maselli},\ and\ \citenamefont
  {Cardoso}}]{Figueiredo:2023gas}%
  \BibitemOpen
  \bibfield  {author} {\bibinfo {author} {\bibfnamefont {E.}~\bibnamefont
  {Figueiredo}}, \bibinfo {author} {\bibfnamefont {A.}~\bibnamefont {Maselli}},
  \ and\ \bibinfo {author} {\bibfnamefont {V.}~\bibnamefont {Cardoso}},\ }\href
  {\doibase 10.1103/PhysRevD.107.104033} {\bibfield  {journal} {\bibinfo
  {journal} {Phys. Rev. D}\ }\textbf {\bibinfo {volume} {107}},\ \bibinfo
  {pages} {104033} (\bibinfo {year} {2023})},\ \Eprint
  {http://arxiv.org/abs/2303.08183} {arXiv:2303.08183 [gr-qc]} \BibitemShut
  {NoStop}%
\bibitem [{\citenamefont {Konoplya}\ \emph
  {et~al.}(2025{\natexlab{d}})\citenamefont {Konoplya}, \citenamefont
  {Khrabustovskyi}, \citenamefont {Kříž},\ and\ \citenamefont
  {Zhidenko}}]{Konoplya:2025mvj}%
  \BibitemOpen
  \bibfield  {author} {\bibinfo {author} {\bibfnamefont {R.~A.}\ \bibnamefont
  {Konoplya}}, \bibinfo {author} {\bibfnamefont {A.}~\bibnamefont
  {Khrabustovskyi}}, \bibinfo {author} {\bibfnamefont {J.}~\bibnamefont
  {Kříž}}, \ and\ \bibinfo {author} {\bibfnamefont {A.}~\bibnamefont
  {Zhidenko}},\ }\href {\doibase 10.1088/1475-7516/2025/04/062} {\bibfield
  {journal} {\bibinfo  {journal} {JCAP}\ }\textbf {\bibinfo {volume} {04}},\
  \bibinfo {pages} {062} (\bibinfo {year} {2025}{\natexlab{d}})},\ \Eprint
  {http://arxiv.org/abs/2501.16134} {arXiv:2501.16134 [gr-qc]} \BibitemShut
  {NoStop}%
\bibitem [{\citenamefont {Chowdhury}\ \emph {et~al.}(2025)\citenamefont
  {Chowdhury}, \citenamefont {Sen}, \citenamefont {Chakrabarti},\ and\
  \citenamefont {Das}}]{Chowdhury:2025tpt}%
  \BibitemOpen
  \bibfield  {author} {\bibinfo {author} {\bibfnamefont {A.}~\bibnamefont
  {Chowdhury}}, \bibinfo {author} {\bibfnamefont {G.}~\bibnamefont {Sen}},
  \bibinfo {author} {\bibfnamefont {S.}~\bibnamefont {Chakrabarti}}, \ and\
  \bibinfo {author} {\bibfnamefont {S.}~\bibnamefont {Das}},\ }\href@noop {} {\
   (\bibinfo {year} {2025})},\ \Eprint {http://arxiv.org/abs/2503.08528}
  {arXiv:2503.08528 [gr-qc]} \BibitemShut {NoStop}%
\bibitem [{\citenamefont {Heydari-Fard}\ \emph {et~al.}(2025)\citenamefont
  {Heydari-Fard}, \citenamefont {Heydari-Fard},\ and\ \citenamefont
  {Riazi}}]{Heydari-Fard:2024wgu}%
  \BibitemOpen
  \bibfield  {author} {\bibinfo {author} {\bibfnamefont {M.}~\bibnamefont
  {Heydari-Fard}}, \bibinfo {author} {\bibfnamefont {M.}~\bibnamefont
  {Heydari-Fard}}, \ and\ \bibinfo {author} {\bibfnamefont {N.}~\bibnamefont
  {Riazi}},\ }\href {\doibase 10.1007/s10714-025-03382-5} {\bibfield  {journal}
  {\bibinfo  {journal} {Gen. Rel. Grav.}\ }\textbf {\bibinfo {volume} {57}},\
  \bibinfo {pages} {49} (\bibinfo {year} {2025})},\ \Eprint
  {http://arxiv.org/abs/2408.16020} {arXiv:2408.16020 [gr-qc]} \BibitemShut
  {NoStop}%
\end{thebibliography}%

\end{document}